\documentclass[pra,a4paper,twocolumn,superscriptaddress,longbibliography,nofootinbib]{revtex4-2}
\usepackage[utf8]{inputenc}
\usepackage{amssymb}
\usepackage{amsmath}
\usepackage{amsfonts}
\usepackage{amsthm}
\usepackage{amsbsy}
\usepackage{bm}
\usepackage{bbold}

\usepackage{graphicx}
\usepackage[dvipsnames]{xcolor}
\usepackage{multirow}
\usepackage{float}
\usepackage{comment}
\usepackage{ulem}
\usepackage{relsize}
\usepackage{cmap}
\usepackage{physics}
\usepackage{cancel}
\usepackage[colorlinks]{hyperref}

\DeclareMathOperator{\sgn}{sgn}

\DeclareMathOperator{\re}{Re}

\DeclareMathOperator{\spp}{sp}

\newcommand{\NLSM}{{NL$\sigma$M}}
\newcommand{\iqHe}{{iqHe}}
\newcommand{\sqHe}{{sqHe}}

\begin{document}
	
	\title{The class C quantum network model with random tunneling and its nonlinear sigma model representation}

\author{D. S. Katkov}

\affiliation{Moscow Institute for Physics and Technology, 141700, Moscow, Russia}

\affiliation{\mbox{L. D. Landau Institute for Theoretical Physics, Semenova 1-a, 142432, Chernogolovka, Russia}}
    
       \author{M. V. Parfenov}

\affiliation{\mbox{L. D. Landau Institute for Theoretical Physics, Semenova 1-a, 142432, Chernogolovka, Russia}}
       
        \affiliation{Department of Physics, HSE University, 101000 Moscow, Russia}

\affiliation{Laboratory for Condensed Matter Physics, HSE University, 101000 Moscow, Russia}

\author{I. S. Burmistrov}

\affiliation{\mbox{L. D. Landau Institute for Theoretical Physics, Semenova 1-a, 142432, Chernogolovka, Russia}}

\affiliation{Laboratory for Condensed Matter Physics, HSE University, 101000 Moscow, Russia}

	\date{\today \, v4} 
	
\begin{abstract}
The spin quantum Hall effect  is a relative of the integer quantum Hall effect, characterized by integer quantized spin Hall conductance. In this work, we formulate and investigate a quantum network model consisting of $\textsf{N}$ channels per chiral link, preserving the fundamental symmetries of the spin quantum Hall effect. We demonstrate that, in the general case, the triplet sector of the theory remains coupled to the singlet sector. In the large-$\textsf{N}$ limit, we systematically derive the effective long-distance, low-energy field theory, identified as a nonlinear sigma model. Our analysis reveals that while triplet modes are typically massive and do not influence the large-$\textsf{N}$ nonlinear sigma model, specific conditions exist where these modes become ``soft'', thereby increasing the ultraviolet cutoff length of the effective theory. Furthermore, by calculating the bare longitudinal and spin Hall conductances, we show that the standard saddle-point approximation fails in regimes with significant tunneling asymmetry between even and odd links. Finally, we establish that the introduction of a Zeeman field not only breaks the 
SU(2)
 symmetry of the nonlinear sigma model action but also generates a term that explicitly violates inversion symmetry.
\end{abstract}	
    
	\maketitle

\section{Introduction}

Anderson localization \cite{Anderson58} in a system of non-interacting fermions in a random potential  has been attracting researchers for more than 65 years (see Refs. \cite{Evers2008,Ryu2016} for reviews).  
As is known \cite{Wigner1951,Dyson1962a,Dyson1962b,Zirnbauer1996,Zirnbauer1997,Zirnbauer2005}, there exist ten different (Altland-Zirnbauer) symmetry classes of disordered non-interacting Hamiltonians. 
In each spatial dimension, there are five of these ten symmetry classes with nontrivial topology that affects physics 
\cite{Schnyder2008,Schnyder2009,Kitaev2009}. A well-known example of such a 
situation is the integer quantum Hall effect ({\iqHe}) \cite{Klitzing1980,Tsui1981}. 
Localization effects in each of the ten symmetry classes can be studied using the effective low-energy field theory -- the nonlinear sigma model ({\NLSM}) (see Ref. \cite{Evers2008} for a review). 
The corresponding {\NLSM} may include the topological term (either the theta-term or the Wess-Zumino-Witten-Novikov term). 
These terms only arise in symmetry classes with non-trivial topology, and are essential for understanding deviations from the conventional picture of Anderson localization.

In two spatial dimensions (2D), a topological Anderson transition is typically a strong coupling phenomenon that may not be accessible within the {\NLSM}. For example, in the case of the {\iqHe}, a critical theory for the plateau-to-plateau transitions is not known and remains a subject of debate in the literature \cite{Zirnbauer1999,Kettemann1999,Bhaseen2000,Tsvelik2001,Tsvelik2007,Bondesan2017,Zirnbauer2019,Zirnbauer2021,Zirnbauer2024}. On the other hand, assuming the local conformal invariance and Abelian fusion rules for pure scaling field theory operators, their scaling exponents at the {\iqHe} criticality have been proven to have a parabolic form with only one free parameter \cite{Bondesan2017,Karcher2021,Padayasi2023}. However, numerical simulations do not support this
prediction \cite{Obuse2008,Evers2008,Karcher2021}. Recent studies have found that the variation of the localization length exponent in the {\iqHe} transition depends on the geometry of the random potential \cite{Sedrakyan2017,Sedrakyan2019,Sedrakyan2021,Dresselhaus2022,Topchyan2024,Gruzberg2026}. 
Despite these controversies regarding critical theory, 
the non-perturbative analysis of {\NLSM} with the topological theta-term \cite{Levine1983,pruisken1984localization} allows us to understand the structure of the phase diagram \cite{Khmelnitskii1983,Pruisken1985}, explain the integer-valued quantization of the Hall conductance \cite{pruisken1987quasiparticles,pruisken1987quasiparticlesB,pruisken1995cracking,Pruisken2005,Pruisken2007}, and develop a coherent picture of {\iqHe} consistent with experimental findings \cite{Pruisken1988,Koch1991,Pruisken2000,Li2005,Pruisken2006,Li2009, Li2010,Shayegan2023,Kaur2024,Yeh2024}.  

In some ways, the opposite situation exists with the spin quantum Hall effect ({\sqHe}), which is a close relative of the {\iqHe}. The {\sqHe} occurs in the unitary superconducting class C \cite{Volovik1997,Kagolovsky1999,Senthil1999}.  
The response of the spin current to the gradient of the external magnetic field is characterized by the integer quantized spin Hall conductivity.~\footnote{Similar relation between the spin current and the gradient of the magnetic field is realized in thin films of  superfluid {${}^3$He-A}~\cite{GEVolovik_1989,GEVolovik_1989_2}.} In contrast to the {\iqHe}, the strong coupling regime of the {\sqHe} is partially accessible via a mapping to the classical percolation problem. In particular, 
the critical value of the spin conductance, the critical exponent of the localization length, and an infinite set of anomalous dimensions of local operators were computed analytically and the obtained results are in agreement with numerical data
\cite{Evers1997,Gruzberg1999,Cardy2000,Beamond2002,Mirlin2003,Evers2003,Subramaniam2008,Puschmann2021,Karcher2022,Karcher2022b,Karcher2023a,Karcher2023}. These analytical findings indicate a lack of local conformal symmetry at the {\sqHe} transition, which complicates the search for a critical theory describing the {\sqHe} transition. A lack of a critical theory for the {\sqHe} transition (similar to the {\iqHe} transition) makes non-perturbative weak-coupling analysis of the corresponding {\NLSM} relevant. In contrast to the {\iqHe}, such non-perturbative analysis of the replica {\NLSM} has  only recently been started \cite{Parfenov2024,Parfenov2025}. 

Our current understanding of the critical exponents that characterize the {\iqHe} transitions is based on the Chalker-Coddington network model \cite{Chalker1988} (see Refs. \cite{Evers2008,Slevin2012T}  for a review). It has been shown \cite{Zirnbauer1997M} that
this model can be mapped to the supersymmetric {\NLSM}. Recently, the SU(2) extension of the Chalker-Coddington network model which belongs to the symmetry class C has been mapped to a supersymmetric {\NLSM} for the class C \cite{Davis2019}. Chalker-Coddington network model has non-random amplitudes for the tunneling between the links. For the symmetry class A, the quantum network model with a random tunneling between the links has been proposed and mapped to {\NLSM} in the continuum limit \cite{DHLee1998}. However, there exists no similar mapping to {\NLSM} for the quantum network model with the symmetry class C. 

In this work, we generalize the quantum network model, which was proposed for the {\sqHe} in Ref. \cite{Senthil1999}, by extending each link to have $\textsf{N}$-channels 
and by considering the most general random tunneling consistent with the symmetry class C. We demonstrate that the continuum limit of this model at large $\textsf{N}$ is mapped to the {\NLSM} in a weak coupling metallic regime. We find that, in general, the triplet sector of the theory is not decoupled from the singlet sector. Typically, the modes in the triplet sector are massive and, thus, do not affect the {\NLSM} action at large $\textsf{N}$. That is, the picture of Anderson localization/delocalization transitions in the infrared limit remains unchanged. However, there exist special cases where the triplet modes become ``soft'' and, as a result, increase the ultraviolet cut-off length for the {\NLSM}. By deriving the expressions for the bare spin longitudinal and Hall conductance, we find that in cases with relatively strong asymmetry in tunneling between even and odd links, the standard saddle point approximation breaks down. We also find that the presence of a Zeeman field not only breaks the SU(2) symmetry of the {\NLSM} action, but also leads to a term that breaks inversion symmetry.

The outline of the paper is as follows. In Sec. \ref{Sec:2} we formulate the quantum network model with $\textsf{N}$-channels in each link. The averaging over random tunneling is discussed in Sec. \ref{Sec:3}. The derivation of {\NLSM} is given in Sec. \ref{Sec4}. In Sec. \ref{Sec5} we analyze the effect of the triplet modes on the {\NLSM} action. We end the paper with discussions and conclusions in Sec. \ref{Sec6}. Some technical details are presented in Appendix.

\begin{figure}[t]
\centerline{\includegraphics[width=0.95\columnwidth]{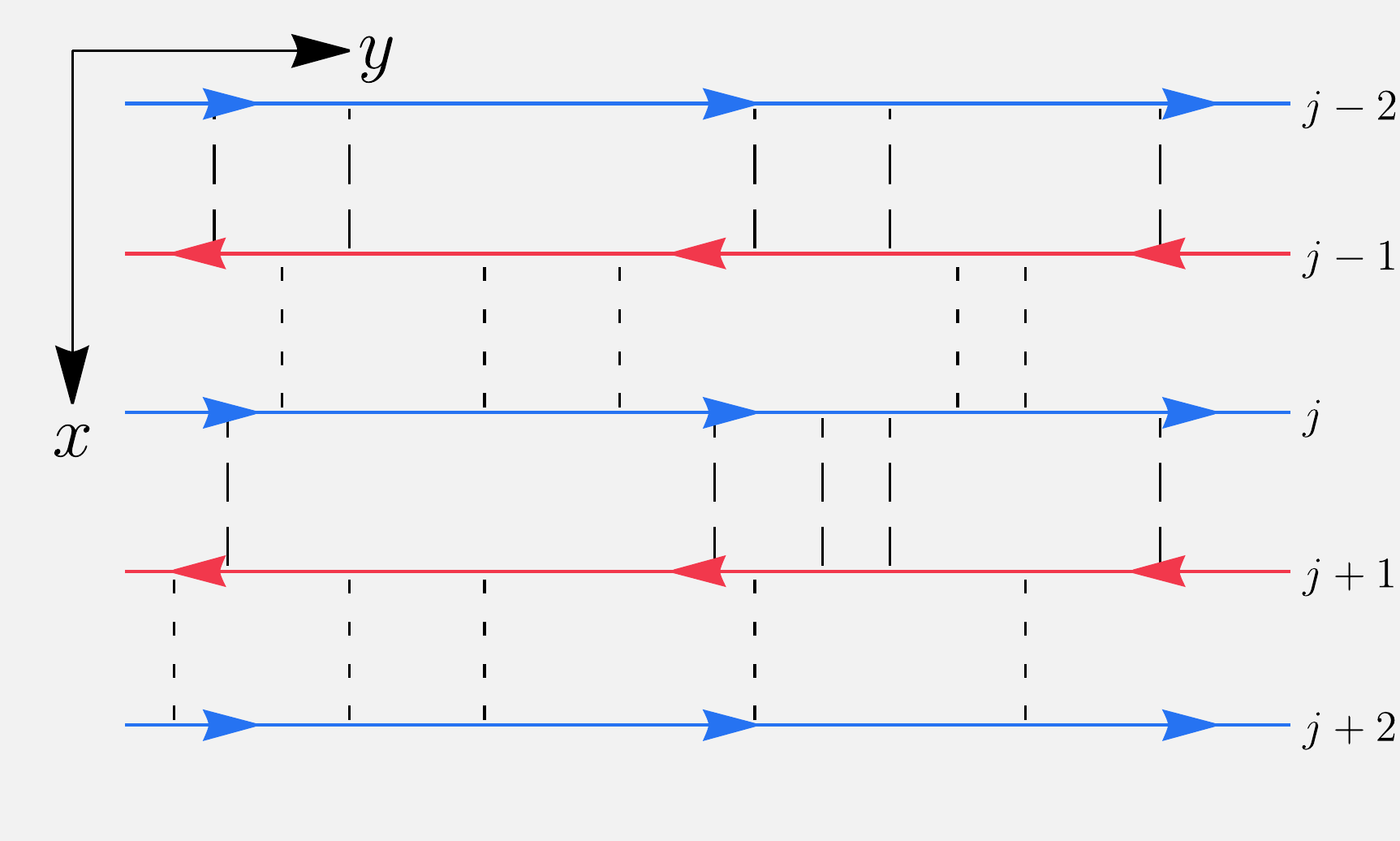}}
\caption{Sketch of the quantum network model. Red and blue lines indicates one-dimensional chiral links, dashed lines corresponds to random tunneling between links.}
\label{fig:model_sheme}
\end{figure}

\section{Model \label{Sec:2}}

\subsection{Formulation}

We consider fermions on a two-dimensional quantum network that belongs to symmetry class C (see Fig. \ref{fig:model_sheme}). At the microscopic level, the system is composed of one-dimensional channels (links) labeled by an integer index $j = 1, \dots, M$. The fermions' velocity alternates in sign on even and odd links, reflecting the chiral motion induced by a perpendicular magnetic field. We encode this by defining the velocity on the $j$-th link as $v_j = (-1)^j v$. Each link hosts $\textsf{N}$ species of spin-$1/2$ fermions. In order to distinguish them we introduce a flavor index $\textsf{f}=1,\dots,\textsf{N}$.

Fermions are subject to  random scattering. To address this, we employ the replica method and introduce $N_r$ copies of the system with the same realization of the random disorder. The fermions in each copy are designated by a replica index $\alpha=1,\dots,N_r$. The action in imaginary time ($\beta=1/T$ where $T$ is the temperature) reads
\begin{gather}
\mathcal{S} = 
\int\limits_0^\beta d\tau \int dy 
\Bigl [ 
\sum_{\sigma=\uparrow,\downarrow} \bar{\psi}_\sigma(i \hat V \partial_y -\partial_\tau - \hat \Sigma_0) \psi_\sigma
 - \bar{\psi}_\uparrow \hat \Sigma_+ \bar{\psi}^T_\downarrow \notag \\ - 
\psi^T_\downarrow \hat \Sigma_- \psi_\uparrow - \bar{\psi}_\uparrow \hat \Sigma_3^{(Z)}\psi_\uparrow + \bar{\psi}_\downarrow \hat \Sigma_3^{(Z)}\psi_\downarrow \Bigr ] .
\label{eq:1}
\end{gather}
Here $\bar{\psi}_\sigma=\{\bar{\psi}_{\sigma,\alpha,j,\textsf{f}}\}$ and $\psi_\sigma=\{\psi_{\sigma,\alpha,j,\textsf{f}}\}^T$ are Grassmann fields corresponding to the creation and annihilation fermionic field operators. The velocity matrix is given by 
$\hat V = 1_{N_r}\otimes \hat v \otimes 1_{\textsf{N}}$, where 
$\hat v_{jj^\prime}{=}v_j \delta_{jj^\prime}$. Four matrices $\hat \Sigma_{0,\pm}$ and $\hat \Sigma_{3}^{(Z)}$ act in the replica, link, and flavor spaces and  are defined as follows
\begin{gather}
\hat\Sigma_0  = 1_{N_r}\otimes (\hat \eta^{(3)}+i \hat \eta^{(0)}), \quad  
\hat \Sigma_{\pm}  =  1_{N_r}\otimes (\hat \eta^{(1)}\mp i \hat \eta^{(2)}) , \notag \\
\hat\Sigma_3^{(Z)}  = 1_{N_r}\otimes \hat{b} \otimes 1_{\textsf{N}} .
\end{gather}
Here the real symmetric matrices 
\begin{equation}
\eta^{(c)}=\eta^{(c)T}=\eta^{(c)\dag},\quad  c=1,2,3    
\end{equation}
and the real skew-symmetric matrix
\begin{equation}
\eta^{(0)}=-\eta^{(0)T}=-\eta^{(0)\dag}    
\end{equation}
 operate in the combined link and flavor spaces. They describe random scattering of fermions. The matrix $\hat b_{jj^\prime}=\mu_B B_j \delta_{jj^\prime}$ describes the Zeeman term with a spatially dependent magnetic field $B_j$.

\subsection{Symmetries}

The Hamiltonian \eqref{eq:1} contains terms with two annihilation or creation fermionic operators. Thus, the U(1) charge symmetry is broken in the Hamiltonian \eqref{eq:1}. As is well known in the context of superconductivity, it is convenient to double the variables. Also, for convenience, we perform a transformation of fermionic fields from the imaginary time to the fermionic Matsubara frequencies, $\varepsilon_n=\pi T(2n+1)$. It implies that fermionic fields $\bar{\psi}$ and $\psi$ acquire additional index $\varepsilon_n$. Thus we introduce 
\begin{gather}
\Xi=\frac{1}{\sqrt{2}}\begin{pmatrix}\psi_\uparrow\\L_0\bar{\psi}^T_\downarrow\\-i\psi_\downarrow\\iL_0\bar{\psi}_\uparrow^T\end{pmatrix},\, 
 \overline{\Xi} = \frac{1}{\sqrt{2}}\begin{pmatrix}\bar{\psi}_\uparrow , \psi^T_\downarrow L_0, i\bar{\psi}_\downarrow , - i\psi^T_\uparrow L_0 \end{pmatrix} ,
\end{gather}
where the matrix 
\begin{equation}
  L_0 = \ell_0\otimes 1_{N_r}\otimes 1_{M} \otimes 1_{\textsf{N}}, \qquad (\ell_0)_{\varepsilon_n,\varepsilon_m}=\delta_{\varepsilon_n,-\varepsilon_{m}} .
\end{equation}
We note that the fields $\overline{\Xi}$ and $\Xi$ are not independent. They are related by the linear transformation 
\begin{equation}
\overline{\Xi} = \Xi^T \hat{C}
\label{eq:rel:Xi:Xi}
\end{equation}
with the charge conjugation matrix 
\begin{gather}
 \hat{C} = L_0 \otimes (i s_2) \otimes \sigma_2 = \begin{pmatrix}
0 & L_0 \otimes \sigma_2 \\
-L_0 \otimes \sigma_2 & 0 
    \end{pmatrix}, \notag \\
    \hat{C}^2=-1 , \qquad \hat{C}^T=\hat{C} .
\end{gather}
Here we introduce two sets of standard Pauli matrices $\sigma_c$ and $s_c$ with $c=0,1,2,3$. 
Following standard convention, we consider that the matrices $\sigma_c$ act on the spin space while the matrices $s_c$
act on the particle-hole (Nambu) space. 

Next we rewrite the action~\eqref{eq:1} as follows
\begin{align}
  \mathcal{S} & = 
\int dy \,\bar{\Xi} \, \mathcal{H} \,
\Xi  , \notag \\
\mathcal{H} & = i \check V \partial_y +i \check{\varepsilon} - i \check{\Upsilon}_0 - \sum_{c=1}^3 \check{\Upsilon}_c - \check{\Upsilon}^{(Z)} ,
\label{eq:1:doubled}
\end{align}  
where $\check V=1_{2N_m} \otimes \hat{V} \otimes s_0 \otimes \sigma_0$. We also introduce the following matrices of size $2N_m \times N_r\times M\times \textsf{N}\times 4$ ($2N_m$ is the number of Matsubara frequencies involved):
\begin{equation}
\begin{split}
 \check{\Upsilon}_c & = 1_{2N_m} \otimes 1_{N_r}\otimes \hat{\eta}^{(c)}\otimes s_0\otimes \sigma_c ,  \\
 \check{\Upsilon}^{(Z)} & = 1_{2N_m}  \otimes 1_{N_r}\otimes \hat{b}\otimes s_3 \otimes \sigma_0 , \\
 \check{\varepsilon} & = \hat \varepsilon \otimes 1_{N_r}\otimes 1_{M} \otimes 1_{\textsf{N}} \otimes s_0\otimes \sigma_0 .
 \end{split}
\end{equation}
Here $\hat\varepsilon_{\varepsilon_n,\varepsilon_m}=\varepsilon_n \delta_{\varepsilon_n,\varepsilon_m}$ is the Matsubara frequency matrix. We emphasize that the relation \eqref{eq:rel:Xi:Xi} is nothing but a realization of the 
antiunitary Bogoliubov-de Gennes (BdG) (charge-conjugation) symmetry \cite{EversMirlin}. Indeed, the Hamiltonian in Eq. \eqref{eq:1:doubled} satisfies the symmetry relation $\mathcal{H}^T=\hat{C} \mathcal{H} \hat{C}$.

In the absence of the Zeeman splitting, $\check{\Upsilon}^{(Z)}=0$, the action \eqref{eq:1:doubled} involves only the matrix $s_0$. It indicates that the action is invariant under the global SU(2) rotation in the Nambu space: $\Xi\to\hat U \Xi$ and $\bar{\Xi}\to\bar{\Xi} \hat U^\dag$, where $\hat U = 1_{2N_m} \otimes 1_{N_r} \otimes 1_M \otimes 1_{\textsf{N}}\otimes \hat{u}\otimes \sigma_0$ with $\hat{u}\in SU(2)$.~\footnote{We note that, one has to preserve the relation \eqref{eq:rel:Xi:Xi} between $\bar{\Xi}$ and $\Xi$. Thus the relation $\hat u=s_2 \hat u^* s_2$ should hold. Indeed, it satisfies for $\hat{u}\in SU(2)$.} The Zeeman term explicitly breaks  the SU(2) spin rotational symmetry down to $U(1)$, i.e., the action \eqref{eq:1:doubled} is invariant under the global rotation with $\hat{u}= \exp(i \theta s_3)$.

\subsection{Random tunneling}

To formulate a specific microscopic model, we assume that the random tunneling of fermions occurs only between neighboring links. Additionally, during the tunneling process, the flavor index of the particles is randomly changed. Therefore, we introduce the following parametrization for the matrices $\hat\eta^{(c)}$ and $\hat\eta^{(0)}$: 
\begin{gather}
\hat\eta^{(c)}_{j\textsf{f};j^\prime \textsf{f}^\prime} = t_j^{(c)}\nu^{(c)}_{\textsf{f}\textsf{f}'} \delta_{j^\prime,j+1}+t_{j^\prime}^{(c)} \nu^{(c)}_{\textsf{f}'\textsf{f}}\delta_{j,j^\prime+1}
+\mu^{(c)}_j \phi^{(c)}_{\textsf{f}\textsf{f}'}\delta_{jj^\prime} , \notag \\
\hat\eta^{(0)}_{j\textsf{f};j'\textsf{f}^\prime} = t_j^{(0)} \nu^{(0)}_{\textsf{f}\textsf{f}'}\delta_{j^\prime,j+1}-t_{j^\prime}^{(0)} \nu^{(0)}_{\textsf{f}'\textsf{f}}\delta_{j,j^\prime+1} .
\label{eta:eq:1}
\end{gather}
Here $t_j^{(0)}(y)$, $t_j^{(c)}(y)$, and $\mu_j^{(c)}(y)$  are independent uncorrelated real random Gaussian variables with zero mean and the following variances 
\begin{equation}
\begin{split}
\langle t_j^{(0)}(y)t_{j'}^{(0)}(y')\rangle  & =\delta_{jj'}\Delta_j\delta(y-y') ,\\
\langle t_j^{(c)}(y)t_{j'}^{(c^\prime)}(y')\rangle  & =\delta_{cc^\prime}\delta_{jj'}\gamma_j\delta(y-y') , \\
\langle \mu_j^{(c)}(y)\mu_{j'}^{(c')}(y')\rangle& =\delta_{cc'}\delta_{jj'}m_j\delta(y-y') . 
\end{split}
\label{eq:t:av}
\end{equation}
For reasons to be explained shortly, in what follows we assume that $\Delta_j, \gamma_j, m_j >0$.
The flavor-space matrices $\nu^{(a)}$ and $\phi^{(a)}$ are independent real random Gaussian matrices with zero mean. Additionally, we assume that the matrices $\phi^{(c)}$ are symmetric, $\phi^{(c)}=\phi^{(c)T}$. Their variances are given by
\begin{equation}
\begin{split}
\langle \phi^{(c)}_{\textsf{f}\textsf{f}'}\phi^{(c')}_{\textsf{f}_1\textsf{f}_1'} \rangle & = \frac{1}{2}\delta_{cc'} \bigl (\delta_{\textsf{f}\textsf{f}_1}\delta_{\textsf{f}'\textsf{f}_1^\prime}+\delta_{\textsf{f}'\textsf{f}_1}\delta_{\textsf{f}\textsf{f}_1^\prime}\bigr )  ,
\\
\langle \nu^{(c)}_{\textsf{f}\textsf{f}'}\nu^{(c')}_{\textsf{f}_1\textsf{f}_1'} \rangle & = \delta_{cc'} \delta_{\textsf{f}\textsf{f}_1}\delta_{\textsf{f}'\textsf{f}_1^\prime} ,  \\
\langle \nu^{(0)}_{\textsf{f}\textsf{f}'}\nu^{(0)}_{\textsf{f}_1\textsf{f}_1'} \rangle & = \delta_{\textsf{f}\textsf{f}_1}\delta_{\textsf{f}'\textsf{f}_1^\prime} .
\end{split}
\label{eq:phi:av}
\end{equation}

We note that our model for $\textsf{N}=1$ is equivalent to the most general form of a network for the class C introduced in Ref. \cite{Senthil1999}. In particular, we generalize it by introducing an arbitrary number of fermion flavors $\textsf{N}$.

To elucidate the meaning of the variances $\Delta_j$ and $\gamma_j$ it is instructive to consider the $2\times 2$ matrix in spin space 
\begin{equation}
\mathcal{T}_j= i t_j^{(0)}\sigma_0+\sum_{c=1}^3 t_j^{(c)}\sigma_c.
\end{equation}
This matrix has the following eigenvalues 
\begin{equation}
t_j^{(\pm)} = i t_j^{(0)}\pm \sqrt{t_j^{(1)2}+t_j^{(2)2}+t_j^{(3)2}}  .  
\end{equation}
Importantly,  $|t_j^{(+)}|=|t_j^{(-)}|$, which reflects the presence of the spin symmetry. We note that
\begin{equation}
\begin{split}
    \langle|t_j^{(\pm)}|^2\rangle =\lambda_j & = \Delta_j+3\gamma_j, \\ 
 \langle \sin^2(\arg t_j^{(\pm)})\rangle & = \frac{\Delta_j}{(\sqrt{\Delta_j}+\sqrt{\gamma_j})^2} .
 \end{split}
\end{equation}
At $\Delta_j=\gamma_j$ we find $\langle \sin^2(\arg t_\pm)\rangle=1/4$ that corresponds to the uniform distribution of the unit vector in the $4$-dimensional space.

\section{Disorder averaging and ballistic renormalization group\label{Sec:3}}

The derivation of the {\NLSM} requires performing an average over disorder realizations. Using the replica trick, we can average the partition function over disorder. The resulting action then contains terms of fourth order in fermionic operators. Due to this, the action requires so-called ballistic renormalization, which is not related to the effects of diffusive collective modes.

\subsection{Disorder averaged action}

As the first step towards the derivation of the {\NLSM}, we split the action \eqref{eq:1:doubled} into two parts:
\begin{equation}
  \mathcal{S} = \mathcal{S}_0 +\mathcal{S}_{\rm dis}, \quad \mathcal{S}_0 = 
\int dy \,\bar{\Xi}  
\Bigl (i \check V \partial_y +i \check{\varepsilon} - \check{\Upsilon}^{(Z)}\Bigr )\Xi    \label{eq:dis:S00}
\end{equation}
and 
\begin{gather}
 \mathcal{S}_{\rm dis}
 = -\int dy\sum_{j} \Bigl [\overline{\Xi}_{j} \Phi_j \Xi_{j}
+ 2 \overline{\Xi}_{j}\mathcal{N}_j \Xi_{j+1}\Bigr] .
 \label{eq:disbefav1}
\end{gather}
Here we introduce two random matrices:
\begin{align}
\Phi_j &= \sum_{c=1}^3 1_{2N_m}\otimes 1_{N_r}\otimes \phi^{(a)} \otimes s_0\otimes (\mu_j^{(c)} \sigma_c) , \notag \\
    \mathcal{N}_j &= i t_j^{(0)} 1_{2N_m}\otimes 1_{N_r}\otimes \nu^{(0)}\otimes s_0\otimes \sigma_0 \notag \\ & + \sum_{c=1}^3 t_j^{(c)}1_{2N_m}\otimes 1_{N_r} \otimes \nu^{(c)}\otimes s_0\otimes \sigma_c \Bigr ) .
\end{align}
Next we perform averaging of the partition function over the random Gaussian matrices $\Phi_j$ and $\mathcal{N}_j$. The averaging results in the substitution of $\mathcal{S}_{\rm dis}$ by~\footnote{\label{FootnoteFierz}In the derivation of Eq. \eqref{eq:dis:av:00} we used the following Fierz identities involving $\bm{\sigma}=\{\sigma_1,\sigma_2,\sigma_3\}$: $\spp \bm{\sigma} A \bm{\sigma} B {=} 2 \spp A \spp B {-} \spp A B$ and  $\spp \bm{\sigma} A \spp \bm{\sigma} B {=} 2 \spp A B {-} \spp A \spp B$ where the trace operates in spin space.} 
\begin{gather}
\mathcal{S}_{\rm av} = \frac{1}{2}\langle \mathcal{S}_{\rm dis}^2\rangle = - \frac{1}{2} \int dy \sum_{j,j'}  \sum_{c=0}^3 
\beta_{jj'}^{(c)} \tr  D^{(c)}_{j} D^{(c)}_{j'}  ,
\label{eq:dis:av:00}
\end{gather}
where
\begin{equation}
\begin{split}
 \beta^{(0)}_{jj^\prime} & = \frac{3}{2} m_j
\delta_{jj'}+ \lambda_j \delta_{j',j+1}+\lambda_{j'}\delta_{j,j'+1} , \\
 \beta^{(c)}_{jj^\prime} & = \frac{1}{2} m_j
\delta_{jj'}+ (\gamma_j{-}\Delta_j) \delta_{j',j+1}+(\gamma_{j'}{-}\Delta_{j'})\delta_{j,j'+1}.
\end{split}
\label{eq:beta:def1}
\end{equation}
In addition, we introduce the matrices $D^{(c)}_{j}$ operating on the Matsubara, replica, and Nambu spaces. They are formally defined as ($c=0,1,2,3$)
\begin{equation}
    [D^{(c)}_{j}]_{\varepsilon_n,\alpha,s;\varepsilon_m,\beta,s^\prime} = \spp_{\textsf{N}{\times}2} \, \sigma_c\, \Xi_{\varepsilon_n,\alpha,j,s} \overline{\Xi}_{\varepsilon_m,\beta,j,s^\prime} ,
    \label{eq:Def:D}
\end{equation}
where the trace is taken over the spin and flavor spaces. We note that the matrices $D^{(c)}_{j}$ satisfy the following symmetry relations:
\begin{gather}
  D^{(0)}_{j} = - C D^{(0)T}_{j} C, \quad D^{(c)}_{j} = C D^{(c)T}_{j} C, \quad c=1,2,3
  \notag \\
   C = \ell_0\otimes 1_{N_r} \otimes s_2, \quad C^2=1, \quad C^T=-C .
   \label{eq:def:CC}
\end{gather}
The action after averaging has an additional symmetry related to the global SU(2) rotation in the spin space. We mention that this global symmetry can be extended to a local one, $\Xi_j\to u_j \Xi_j$ and  $\bar{\Xi}_j\to \bar{\Xi}_j u_j^\dag$, in the special case $\Delta_j=\gamma_j$. It is this case that was studied in Refs. \cite{Senthil1999,Senthil2000}. Also, we note that for $\Delta_j=\gamma_j$ and $\textsf{N}=1$, the triplet matrices $D^{(c)}_{j}$, where $c=1,2,3$, can be excluded from the action \eqref{eq:dis:av:00}. In the following, we  consider the general case where this local rotational symmetry is absent.

\subsection{Ballistic renormalization group}

Before deriving the {\NLSM} action, we consider the renormalization of the disorder-averaged action, cf. Eqs~\eqref{eq:dis:S00} and \eqref{eq:dis:av:00}, which were obtained in the previous section. We neglect the effect of the Zeeman field for now. As we will demonstrate below, the connection between non-nearest neighbouring links is generated during the ballistic renormalization group flow. This results in the matrix $\beta^{(c)}$ becoming non-tridiagonal at the length scale where diffusive behaviour starts.


As usual, we split the fermionic fields into fast and slow modes in momentum space by introducing a separating momentum scale $q_\Lambda$. The renormalization of the $\beta^{(a)}$ matrix then comes from the squared term in Eq.~\eqref{eq:dis:av:00}, where we average over four of the eight fermionic fields that are fast:
\begin{gather}
\frac{1}{2} \langle \mathcal{S}_{\rm av}^2 \rangle_0 = 
\frac{1}{2} \int dy dy' \sum_{jj'j_1j_1'}\sum_{c,c'} \beta^{(c)}_{jj'}\beta^{(c')}_{j_1j_1'}
\sum_{z_{k}}
[D^{(c)}_{j}(y)]_{z_1;z_2} 
\notag \\
\times 
  [D^{(c')}_{j_1'}(y')]_{z_3;z_4} \left \langle 
 [D^{(c)}_{j'}(y)]_{z_2;z_1} 
  [D^{(c')}_{j_1}(y')]_{z_4;z_3}
  \right \rangle_0 .\label{eq:RG_no_avg}
\end{gather}
Here $z_k$ is the combined index for the variables $\{\varepsilon_n,\alpha,s\}$. 
We also note that we select the contractions that give contributions proportional to the number of flavors $\textsf{N}$. The omitted contractions yield terms of order of $O\left(\textsf{N}^{0}\right)$ that can be neglected in the large $\textsf{N}\gg 1$ limit. The average $\langle\dots\rangle_0$ is taken with respect to the action $\mathcal{S}_0$, as defined in Eq. \eqref{eq:dis:S00}, with $\check \Upsilon^{(Z)}=0$. By performing the averaging (see Appendix \ref{App:BRG}), we obtain
\begin{gather}
\frac{1}{2} \langle \mathcal{S}_{\rm av}^2 \rangle_{\rm fast} \simeq  
\frac{\textsf{N}}{2\pi v^2}\int\limits_{q_{\Lambda}}^\infty \frac{dq}{q^2} \int dy \sum_{c=0}^3 \sum_{jj'}
(\beta^{(c)}\beta^{(c)})_{jj'} \notag \\
\times \tr  D^{(c)}_{j}(y) D^{(c)}_{j'}(y) .
\label{eq:corr:ballistic}
 \end{gather}
 By introducing dimensionless momentum-dependent disorder strength $\bar{\beta}^{(c)}=\textsf{N} \beta^{(c)}/(\pi v^2 q_\Lambda)$, we obtain the following renormalization group equations
 \begin{equation}
  - \frac{d\bar{\beta}^{(c)}}{d\ln q_\Lambda} = \bar{\beta}^{(c)}- (\bar{\beta}^{(c)})^2  . 
  \label{eq:RG1}
 \end{equation}
As one can see, the singlet ($c=0$) and triplet ($c=1,2,3$) channels are not mixed under renormalization group flow (at least within the one-loop approximation). Next, $\beta^{(c)}=0$ is the unstable fixed point of Eq. \eqref{eq:RG1} at $q_\Lambda \to 0$.  Additionally, even if one starts in the ultraviolet with the matrix $\beta^{(c)}$ which corresponds to  tunneling only between nearest links, see Eq. \eqref{eq:beta:def1}, the nonzero elements connecting non-nearest neighboring links are generated during the renormalization group flow. We note that a similar picture holds for the renormalization of tunneling action at the superconductor--normal metal boundary~\cite{skvorsc2000}. Although, formally, Eq. \eqref{eq:RG1} has a stable fixed point at $\bar{\beta}^{(c)}=1$, it is inaccessible since some element of the matrix $\bar{\beta}^{(c)}$ diverges at a finite length scale. Therefore, generically, $\bar{\beta}^{(c)}$ flows toward some strong coupling fixed point that is beyond the one-loop description.

\section{Derivation of the {\NLSM}\label{Sec4}}

In this section, we perform the derivation of the {\NLSM} action. We do this in several steps. First, we perform the Hubbard-Stratonovich transformation and integrate over fermions. Next, we analyze the saddle point of the theory and the structure of the massive and massless modes. Finally, integrating out the massive modes allows us to obtain the low-energy effective field theory.

\subsection{Hubbard-Stratonovich transformation}

We start by decoupling of the quadratic terms in Eq. \eqref{eq:dis:av:00} by means of the Hubbard-Stratonovich matrix fields $\tilde{Q}_j^{(c)}$, where $c=0,1,2,3$, conjugated to $D_j^{(c)}$. Naturally, the matrix field $\tilde{Q}_j^{(c)}$ acts on the same spaces as $D_j^{(c)}$ and obeys the same symmetry, cf. Eq. \eqref{eq:def:CC}:
\begin{gather}
  \tilde{Q}^{(0)}_{j} = - C \tilde{Q}^{(0)T}_{j} C, \,\, \tilde{Q}^{(c)}_{j} = C \tilde{Q}^{(c)T}_{j} C, \,\, c=1,2,3 .
   \label{eq:def:CC:Q}
\end{gather}
We note the difference in symmetry relations for the singlet, $\tilde{Q}^{(0)}_{j}$, and triplet, $\tilde{Q}^{(1,2,3)}_{j}$ matrix modes. 
After integrating out the fermionic fields $\Xi$ and $\bar{\Xi}$, the action reduces to:
\begin{gather}
\mathcal{S}=\frac{\textsf{N}}{2}\sum_j\Tr\ln\Bigl[(i\tilde{v}_j\partial_y{+}i\tilde\varepsilon{-} \tilde{b}_j)\sigma_0 
{+} i \sum_{j'} \sum_{c=0}^3\beta^{(c)}_{jj'} \tilde{Q}^{(c)}_{j'} \sigma_c \Bigr]
\notag \\
- \frac{1}{2} \int dy\sum_{jj'}\sum_{c=0}^3\beta_{jj'}^{(c)} \tr \tilde{Q}^{(c)}_j \tilde{Q}^{(c)}_{j'} .
\label{action_d=g}
\end{gather}
Here we introduce the following matrices acting in the Matsubara frequency, replica, and Nambu spaces:
\begin{equation}
\begin{split}
        \tilde{v}_j & = v_j\, 1_{2N_m}\otimes 1_{N_r}\otimes s_0,\\
    \tilde{\varepsilon} & = \hat{\varepsilon}\otimes 1_{N_r}\otimes s_0, 
\\
\tilde{b}_j & = \mu_B B_j\, 1_{2N_m}\otimes 1_{N_r}\otimes s_3 .
\end{split}
\end{equation}

As we discussed above, the action \eqref{eq:dis:av:00} has a global SU(2) rotational symmetry in spin space, which can be extended to a local symmetry in the special case with $\Delta_j=\gamma_j$. In this case, $\beta^{(c)}_{jj'}=(m_j/2)\delta_{jj'}$, for $c=1,2,3$, so using Fierz identities in the spin space (see Footnote~\ref{FootnoteFierz}) the action 
\eqref{action_d=g} can be  written solely in terms of the invariant combinations of the  matrix $\hat{Q}_j=\sum_{c=0}^3\tilde Q_{j}^{(c)}\sigma_c$, e.g., $\spp\hat{Q}_j$, $\spp \hat{Q}_j^2$, and $(\spp \hat{Q}_j)^2$, where $\spp$ is the trace over spin space only. Therefore, the action \eqref{action_d=g} becomes invariant under local SU(2) rotations in the spin space for the case  
$\Delta_j=\gamma_j$.

\subsection{The saddle point}

The first line of the action \eqref{action_d=g} is proportional to the number of flavors $\textsf{N}$. Therefore, at $\textsf{N}\gg 1$ we can treat the action \eqref{action_d=g} in the saddle-point approximation.
By varying the action \eqref{action_d=g} (with $\tilde{b}_j=0$ and $\tilde{\varepsilon}=0$) over the matrix field $\tilde{Q}^{(c)}_j$, we obtain the saddle-point equations
\begin{gather}
\sum_{j'}\beta_{jj'}\tilde{Q}^{(c)}_{j'}(y) 
=i \frac{\textsf{N}}{2}\sum_{j'}\beta_{jj'} \tr \sigma_c G_{j'}(y,y) ,
\label{eq:sp:eq}
\end{gather}
where $G_j$ stands for the Green's function corresponding to the inverse operator under the trace-logarithm:
\begin{gather}
G_j^{-1}=i\tilde{v}_js_0  \partial_y 
{+} i \sum_{j'} \sum_{c=0}^3\beta^{(c)}_{jj'} \tilde{Q}^{(c)}_{j'} \sigma_c .
\end{gather}
As one can check, the saddle-point equation has the standard solution
\begin{equation}
\underline{\tilde{Q}}^{(c)}_{j} = \frac{\textsf{N}}{2v} \Lambda \delta_{c0}, \quad \Lambda = \hat\eta \otimes 1_{N_r}\otimes s_0 ,  
\label{eq:sol:eq}
\end{equation}
where $\hat\eta_{\varepsilon_n,\varepsilon_m} = \sgn\varepsilon_n \delta_{\varepsilon_n,\varepsilon_m}$. The corresponding saddle-point Green's function is diagonal in the spin space:
\begin{equation}
    \underline{G}_j^{-1}=\left (i\tilde{v}_j\partial_y  
{+} \frac{i}{2\tau_j}\Lambda\right ) \sigma_0, \qquad \frac{1}{\tau_j} = \frac{\textsf{N}}{v} \sum_{j'}\beta^{(0)}_{jj'}
.
\label{eq:sp:tau}
\end{equation}
We note that there are more solutions to Eq.~\eqref{eq:sp:eq}. We can freely rotate the solution \eqref{eq:sol:eq} in the Matsubara frequency, replica and Nambu spaces with a matrix $t$ independent of the link index $j$ and the spatial coordinate $y$. Then the saddle-point manifold is given by
\begin{equation}
    \underline{\tilde{Q}}^{(c)}_{j} = \frac{\textsf{N}}{2v} t^{-1} \Lambda t \, \delta_{c0}, \qquad C t^{-1}=t^T C .
\label{eq:sspp:11}
\end{equation}
Since the matrix $\tilde{Q}^{(c)}_{j}$ is conjugated (in the sense of the Hubbard-Stratonovich transformation) to the matrix $D^{(c)}_j$, see Eq. \eqref{eq:Def:D}, it is natural to impose the Hermiticity condition on $\underline{\tilde{Q}}^{(0)}_{j}$ to ensure convergence, i.e., to assume that
 $\underline{\tilde{Q}}^{(0)}_{j}=\underline{\tilde{Q}}^{(0)\dag}_{j}$.
 Then together with the symmetry relation \eqref{eq:def:CC:Q}, the saddle-point manifold corresponds to the {\NLSM} target space of class C, $\underline{\tilde{Q}}^{(0)}_{j}\in G/K$ with $G={\rm Sp}(4N_rN_m)$ and $K={\rm U}(2N_rN_m)$.

\subsection{Massive modes}

We start with the triplet spin sector, i.e., the matrix fields  $\tilde{Q}^{(c)}_j$ (with $c=1,2,3$). By expanding the $\tr\ln$ in Eq.~\eqref{action_d=g} to the second order in the matrix fields  $\tilde{Q}^{(c)}_j$ and neglecting spatial derivatives, we obtain the following Gaussian action:
\begin{align}
 \mathcal{S}_{\rm t}^{(2)}  \simeq & -\frac{1}{8} \int dy \sum_{c=1}^3\sum_{jj'}\beta_{jj'}^{(a)}\tr \{\tilde{Q}^{(c)}_j,\Lambda\}\{\tilde{Q}^{(c)}_{j'},\Lambda\} , \notag \\
 & + \frac{1}{8} \int dy \sum_{c=1}^3\sum_{jj'}\tilde{\beta}_{jj'}^{(c)}\tr \bigl [\tilde{Q}^{(c)}_j,\Lambda\bigr ][\tilde{Q}^{(c)}_{j'},\Lambda] ,
 \label{eq:gauss:triple}
 \end{align}
where 
\begin{equation}
   \tilde{\beta}_{jj'}^{(c)} = \beta_{jj'}^{(c)} - \frac{\textsf{N}}{v}\sum_{j''} \beta_{jj''}^{(c)} \tau_{j''}
   \beta_{j''j'}^{(c)} .
\end{equation}
We note that the above expression for $\tilde{\beta}^{(a)}$ is similar to the ballistic renormalization, cf. Eq.~\eqref{eq:corr:ballistic}, computed at the momentum scale corresponding to the inverse mean free path $q_\Lambda \sim 1/(v \tau_j)$. Provided the matrices $\beta^{(c)}$ and $\tilde{\beta}^{(c)}$ have no zero eigenvalues, the matrix fields $\tilde{Q}^{(c)}_j$ (with $c=1,2,3$) are massive and, thus, can be ignored in the course of the derivation of the {\NLSM} action. We will discuss their effect on the {\NLSM} action in Sec. \ref{Sec5} below.

For the singlet sector, a naive expansion in the deviation of the $\tilde{Q}_j^{(0)}$ from its saddle-point value yields exactly the same expressions as in Eq.~\eqref{eq:gauss:triple}, where $\tilde{Q}_j^{(c)}$ is substituted by $\tilde{Q}_j^{(0)}-\underline{\tilde{Q}}_j^{(0)}$. The first line of Eq.~\eqref{eq:gauss:triple} then corresponds to the modes that commute with $\Lambda$. They are massive if the matrix $\beta^{(0)}$ does not have any zero eigenvalues (see Appendix \ref{App_Det1}).~\footnote{We note that there is also the contribution from the Jacobian for the part of $\tilde{Q}_j^{(0)}$ that commutes with $\Lambda$  \cite{Zirnbauer1997}. However, this fact does not change the conclusion regarding the massive modes, which is similar to what happens in the {\iqHe} \cite{pruisken1984localization}.}. 
The second line of Eq.~\eqref{eq:gauss:triple}  describes modes that anti-commute with $\Lambda$. They can be massless, since $\det \tilde{\beta}^{(0)}=0$ by construction. These modes are the ones we are interested in below.

\subsection{The case of strong diagonal disorder $m\gg\gamma_{\pm}, \Delta_{\pm}$ \label{Sec:NLSMder1}}

Although we are interested in deriving the {\NLSM} action for the most general form of the matrices $\beta^{(c)}$, we start with the simplest case. 
Namely, we start from the tridiagonal matrices $\beta^{(c)}$, cf. Eq. \eqref{eq:beta:def1}, with the assumption that there is an even/odd-link asymmetry such that 
$\gamma_{2j}=\gamma_{+}$, $\gamma_{2j+1}=\gamma_{-}$,
$\Delta_{2j}=\Delta_{+}$, and $\Delta_{2j+1}=\Delta_{-}$. In this case, from Eq. \eqref{eq:sp:tau} we find that  the scattering time is independent of the link index: $1/\tau_j=1/\tau=\textsf{N} (3 m+2\lambda_++2\lambda_-)/(2v)$. We also assume that 
$m_j=m\gg\gamma_{\pm}, \Delta_{\pm}$. Under this assumption, as follows from the ballistic renormalization \eqref{eq:corr:ballistic}, the elements of the matrix $\beta^{(a)}$ corresponding to tunneling between non-nearest neighbor links are suppressed. 

We parametrize the singlet matrix field as 
\begin{equation}
\tilde{Q}_j^{(0)}(y) = \frac{\textsf{N}}{2v} t_j^{-1}(y) \Lambda t_j(y) ,
\end{equation}
where $t_j(y)$ has the same symmetries as the matrix $t$ in Eq.~\eqref{eq:sspp:11}. 
Then the effective action \eqref{action_d=g} becomes 
\begin{gather}
\mathcal{S}=\textsf{N}\sum_j\Tr\ln\Bigl[i\tilde{v}_j\partial_y{+}i\tilde\varepsilon{-} \tilde{b}_j 
{+} \frac{i\textsf{N}}{2v} \sum_{j'} \beta^{(0)}_{jj'} t_{j'}^{-1}\Lambda t_{j'} \Bigr]
\notag \\
- \frac{\textsf{N}^2}{8v^2} \int dy\sum_{jj'}\beta_{jj'}^{(0)} \tr t_{j}^{-1}\Lambda t_{j} t_{j'}^{-1}\Lambda t_{j'} .
\label{action_d=g:1}
\end{gather}
Here we neglected all massive modes as discussed above. As usual, we assume that the matrix $t_j$ varies slowly along the link ($y$ coordinate) and changes slowly from link to link (below we denote this direction as $x$). Then, we employ the following expansion up to  second order in spatial gradients
\begin{gather}
t_{j\pm 1}^{-1}\Lambda t_{j\pm 1} \simeq t_{j}^{-1}
\Bigl \{ \Lambda \pm a [L_j^{(x)},\Lambda] 
+ \frac{a^2}{2} [\partial_x L_j^{(x)},\Lambda] \notag \\
+\frac{a^2}{2} [L_j^{(x)},[L_j^{(x)},\Lambda]]
\Bigr \} t_{j} .
\label{eq:der2}
\end{gather}
Here $L^{(x,y)}_j=t_j\partial_{x,y} t_j^{-1}$ and $a$ is the spatial period of the network in real space. By substituting  expression~\eqref{eq:der2} into Eq.~\eqref{action_d=g:1} and expanding $\tr\ln$ upto the second order in the spatial gradients, we obtain
\begin{gather}
\mathcal{S} \simeq - \sum_j\int dy \Biggl \{ \sum_{r=x,y} \pi_{r}^{(0)} \tr [ L_j^{(r)},\Lambda ]^2 - \frac{\textsf{N}}{2}(-1)^j \tr \Lambda L_j^{(y)} \notag \\
+ \frac{\textsf{N}^2 a \tau (\lambda_+-\lambda_-)}{2v}\tr\Lambda[ L_j^{(x)},L_j^{(y)} ]
- \frac{\textsf{N}}{2v}\tr (\tilde{\varepsilon}+i\tilde{b}_j) t_j^{-1}\Lambda t_j \notag \\- \frac{\textsf{N}\tau}{8 v}\tr[ t_j \tilde{b}_j t_j^{-1}, \Lambda]^2
+
\frac{i \textsf{N}\tau}{4v} v_j \tr [t_j \tilde{b}_j t_j^{-1},\Lambda][L_j^{y},\Lambda] 
   \Bigr \} ,
   \label{eq:S:interm}
\end{gather}
where 
\begin{gather}
  \pi_x^{(0)} = \frac{\textsf{N}^2 a^2(\lambda_++\lambda_-)}{16 v^2}\Bigl [ 1{-}\frac{{2}\textsf{N}\tau}{v}\frac{(\lambda_+-\lambda_-)^2}{(\lambda_++\lambda_-)}\Biggr ] ,   \notag \\
  \pi_y^{(0)} = \frac{\textsf{N} v\tau}{8} .
\end{gather}
To proceed further with the action \eqref{eq:S:interm}, we have to resolve three issues. First, we have to express all contributions in terms of the matrix field $Q_j = t_j^{-1}\Lambda t_j$ in order to restore the symmetry of the saddle-point manifold. %
 Second, we have to take the continuum limit, $a\to 0$, explicitly. Third, we have to restore the spatial isotropy, i.e., the symmetry between the $x$ and $y$ directions.   

In order to express the effective action \eqref{eq:S:interm} in terms of the matrix field $Q_j = t_j^{-1}\Lambda t_j$, we use the following identities:
\begin{gather}
\tr [ L_j^{(x)},\Lambda ]^2  =\tr(\partial_{x} Q_j)^2, \, \tr [ L_j^{(y)},\Lambda ]^2  =\tr(\partial_{y} Q_j)^2 , \notag \\
\tr\Lambda [L_j^{(x)},L_j^{(y)}] =-\frac{1}{2}\tr Q_j\partial_x Q_j\partial_y Q_j .
\end{gather}
Next, we define the continuum limit, $a\to 0$, by the following transformation from the sum to the integral,
$a \sum_j \to \int dx$. Also, we rescale the coordinates as $x\to x \zeta \sqrt{\pi_x}$ and $y\to y \zeta \sqrt{\pi_y}$. Here, the parameter $\zeta$  does not appear in the terms involving spatial derivatives. We fix $\zeta$ by requiring that the term proportional to the Matsubara energy has the standard coefficient equal to the density of states (see below).  
Finally, we note that in the continuum limit, the term proportional to $(-1)^j\tr \Lambda L_j^{(y)}$ becomes the edge term and it can then be rewritten as a bulk term:
\begin{gather}
\int dy\sum_j(-1)^j\tr \Lambda L_j^{(y)} = \oint ds \tr \Lambda t \partial_s t^{-1} \notag \\
= \frac{1}{8}\int dx dy \epsilon_{\mu\nu} \tr Q\partial_\mu Q\partial_\nu Q .
\end{gather}
Here  $\epsilon_{\mu\nu}$ stands for the antisymmetric tensor with non-zero elements $\epsilon_{xy}=-\epsilon_{yx}=1$.
We note that the integration contour for $s$ follows the chiral channels (see Fig.~\ref{fig:model_sheme}). This contour encloses only half of the system's area, so an additional factor of $1/2$ appears  when the Stokes' integral is extended to the total area.

In summary, the final form of the {\NLSM} action can be written as:
\begin{align}
\mathcal{S}  &=-\frac{\bar{g}}{16}\int d^2\bm{r} \tr(\nabla Q)^2+\frac{\bar{g}_H}{16}\int d^2\bm{r}\epsilon_{\mu\nu}\tr Q\partial_\mu Q\partial_\nu Q\notag \\
& +\pi \bar{\nu}\int d^2\bm{r}\, \tr \left (\hat{\varepsilon} Q+i\mu_B \overline{\textsf{B}}s_3 Q
+\frac{\tau \mu_B^2 }{4} \overline{\textsf{B}^2}\tr [s_3,Q]^2 \right )
\notag \\
&
-\frac{i \textsf{N}\tau\mu_B}{4} \int d^2\bm{r}\, \overline{\textsf{B}}\, \epsilon_{\mu\nu}\tr s_3 \partial_\mu Q\partial_\nu Q
\notag \\
& 
-\frac{i\mu_B\sqrt{\pi \tau\bar\nu\bar g}}{2v}\int d^2\bm{r}\, \overline{\textsf{B} \textsf{v}}\tr s_3 Q\partial_y Q.
\label{final_action}
\end{align}
The matrix $Q$ satisfies the following constraints:
\begin{gather}
Q^\dag = Q, \quad Q = -C Q^T C, \quad Q^2=1 , \notag \\
Q\in {\rm Sp}(4 N_r N_m)/{\rm U}(2 N_r N_m) .
\end{gather}
Here the matrix $C$ is defined in Eq. \eqref{eq:def:CC}. 
The bare  dimensional spin longitudinal conductance (in units $G_0^s=\hbar/(8\pi)$) is given by 
\begin{gather}
\bar{g} = \textsf{N} \left [\frac{2\textsf{N}\tau}{v} (\lambda_++\lambda_-)\left (1{-}\frac{2\textsf{N}\tau}{v}\frac{(\lambda_+-\lambda_-)^2}{(\lambda_++\lambda_-)}\right )\right ]^{1/2} \notag \\
\simeq 2 \textsf{N} 
\left (\frac{\lambda_++\lambda_-}{3m}\right )^{1/2} .
\label{final_g}
\end{gather}
The bare dimensional spin Hall conductance becomes 
\begin{equation}
 \bar{g}_H = \textsf{N} \left [1+\frac{2\textsf{N}\tau}{v}(\lambda_+-\lambda_-)\right ]   \simeq 
 \textsf{N} \left [1 + \frac{4(\lambda_+-\lambda_-)}{3m}\right ] .
 \label{final_gH}
\end{equation}
We also introduce the bare density of states as
$\bar{\nu} = \textsf{N}/(\pi^2 a v) \equiv \textsf{N}\zeta^2\sqrt{\pi_x\pi_y}/(2 v a{\pi})$. The latter equality fixes the rescaling parameter $\zeta$ as $\zeta^2=2/(\pi\sqrt{\pi_x\pi_y})$.

We note that for $\lambda_+=\lambda_-$, the bare Hall conductance is equal to the integer, $\bar{g}_H = \textsf{N}$. Therefore, for an odd number of flavors the {\NLSM} for the symmetric case $\lambda_+=\lambda_-$ corresponds to the unstable critical line separating phases with the quantized spin Hall conductance in even integers. Interestingly, the bare value of the spin longitudinal conductance is almost insensitive to deviations from the line $\lambda_+=\lambda_-$ by virtue of the assumption $\lambda_\pm \ll m$.     

The quantities $\overline{\textsf{B}}$, $\overline{\textsf{B}^2}$, and $\overline{\textsf{B}\textsf{v}}$ represent the spatial average of the Zeeman magnetic field, its square, and its product with velocity, respectively. In particular, we define
\begin{equation}
\overline{\textsf{B}^m} = \frac{1}{M}\sum_j B^m_j, \qquad  \overline{\textsf{B}\textsf{v}} = \frac{1}{M}\sum_j B_j v_j .
\end{equation}
All four terms in Eq. \eqref{final_action} that involve the Zeeman magnetic field  explicitly break the SU(2) symmetry of the {\NLSM} action. 
In the case of alternating sign of the Zeeman field at even/odd links, $B_j=(-1)^j B$, the average is zero, $\overline{\textsf{B}}=0$. In such a situation, the last term in Eq.~\eqref{final_action} is responsible for the breaking of the SU(2) symmetry. In addition, the last term in Eq. \eqref{final_action} breaks the inversion symmetry if the Zeeman splitting is synchronized with the velocities $v_j$, i.e., $\overline{\textsf{B}\textsf{v}}\neq 0$. This happens, for example, in the case $B_j=(-1)^j B$.

\subsection{The symmetric case \label{Sec:NLSMder2}}

In this section, we derive the {\NLSM} for the symmetric case, where the scattering nodes on the even and odd links are identical. For the model \eqref{eq:beta:def1} the symmetric case corresponds to the choice
$\lambda_+=\lambda_-$. Since, according to the ballistic renormalization group, see Eq. \eqref{eq:RG1}, tunneling between nearest-neighbor links leads to tunneling between distant links, we consider an arbitrary symmetric Toeplitz-type matrix $\beta_{jj'}^{(0)}$, which depends only on the distance $|j-j'|$. 
As before, we ignore all massive modes in this derivation.
 
 In the following, we focus on the infinite number of channels, $M\to \infty$. Therefore, it is convenient to consider the matrix $\beta_{jj'}^{(0)}$ to be of infinite size and define its Fourier transform as 
\begin{equation}
\beta_{jj'}^{(0)} = \int\limits_{-\pi}^\pi \frac{dk}{2\pi}\underline{\beta}(k) e^{i k (j-j')} .
\label{eq:FT:T}
\end{equation}
Introducing new matrix variable $\tilde{Q}_j=\sum_{j'} \beta^{(0)}_{jj'} \tilde{Q}^{(0)}_{j'}$, we rewrite the action \eqref{action_d=g} as 
\begin{align}
\mathcal{S} & =\textsf{N}\sum_j\Tr\ln\Bigl[i\tilde{v}_j\partial_y{+}i\tilde\varepsilon{-} \tilde{b}_j 
{+} i \tilde{Q}_{j}\Bigr] \notag \\
& - \frac{1}{2} \int dy\sum_{jj'}[\beta^{(0)}]^{-1}_{jj'} \tr \tilde{Q}_j \tilde{Q}_{j'} .
\label{action_d=g:002}
\end{align}
We assume that the function $\underline{\beta}(k)$ has no zeroes, such that the matrix $\beta_{jj'}^{(0)}$ is invertible. The saddle-point solution for the matrix $\tilde{Q}_{j}$ becomes
\begin{equation}
  \underline{\tilde{Q}}_j = \frac{1}{2\tau} t^{-1}\Lambda t, \qquad \frac{1}{\tau} =\frac{\textsf{N}}{v}\sum_{j'} \beta^{(0)}_{jj'} = \frac{\textsf{N}}{v}\underline{\beta}(0) ,   
  \label{eq:SP:1}
\end{equation}
where similar to Eq. \eqref{eq:sspp:11}, the matrix $t$ is independent of $y$. Making the rotation $t$ spatially and link dependent, we perform the gradient expansion of the action as before. Hence, we obtain 
\begin{gather}
\mathcal{S} \simeq - \sum_j\int dy \Bigr \{ \pi^{(s)}_{y} \tr [ L_j^{(y)},\Lambda ]^2
+ \pi^{(s)}_x \tr (\partial_x Q_j)^2 
\notag \\
- \frac{\textsf{N}}{2}(-1)^j \tr \Lambda L_j^{(y)} 
- \frac{\textsf{N}}{2v}\tr (\tilde{\varepsilon}+i\tilde{b}_j) Q_j \notag \\- \frac{\textsf{N}\tau}{8 v}\tr[ t_j \tilde{b}_j t_j^{-1}, \Lambda]^2
   \Bigr \} ,
   \label{eq:S:interm:00}
\end{gather}
where $Q_j = t_j^{-1}\Lambda t_j$ and 
\begin{equation}
\pi_x^{(s)} = -\frac{a^2\underline{\beta}''(0)}{16\tau^2\underline{\beta}^2(0)},\qquad 
\pi_y^{(s)} = \frac{\textsf{N} v\tau}{8} .
\end{equation}
Next, repeating exactly the same steps as in Sec.~\ref{Sec:NLSMder1}, we obtain the effective action in the form of Eq.~\eqref{final_action} but with 
\begin{equation}
 \bar{g} = \textsf{N} \left (\frac{2 |\underline{\beta}''(0)|}{\underline{\beta}(0)}\right )^{1/2}, \qquad \bar{g}_H = \textsf{N} .
 \label{eq:ff:g:gH}
\end{equation}
We note that in the case  $\underline{\beta}(k)=3m/2+2\lambda \cos k$, where $\lambda=\lambda_+=\lambda_-$ with $\lambda\ll m$, the results \eqref{eq:ff:g:gH}
reproduce the expressions \eqref{final_g} and \eqref{final_gH}.

\subsection{Asymmetric long-range tunneling \label{Sec:NLSMder3}}

In this section, we allow for both long-range tunneling and an asymmetry between the nodes on the odd and even links. In this case, $\beta_{jj'}^{(0)}$ depends not only on the distance $|j-j'|$ but also on the parity of the $j$-th link.   
We can write the link index $j$ as $j=2n+s$, where $n$ enumerates links on two sublattices which are labeled by $s=0,1$ for the even/odd link, respectively. 
Then, for $j=2n+s$ and $j'=2n'+s'$, we can write $\beta_{jj'}^{(0)}\equiv\beta_{ss'}^{(0)}(n-n')$. Due to the symmetry of the tunneling matrix, $\beta_{jj'}^{(0)}=\beta_{j'j}^{(0)}$, we have $\beta_{ss'}^{(0)}(n-n')=\beta_{s's}^{(0)}(n'-n)$. To make the scattering time $\tau$ independent of the sublattice index $s$, we also assume that $\beta_{00}^{(0)}(n)=\beta_{11}^{(0)}(n)$ for all $n$. With these assumptions, introducing the matrix field $\tilde{Q}_j=\sum_{j'} \beta^{(0)}_{jj'} \tilde{Q}^{(0)}_{j'}$ brings us back to the action \eqref{action_d=g:002}. The saddle-point solution remains unchanged. For an arbitrary integer displacement $r$ we perform a gradient expansion, assuming that the matrix field varies slowly in the $x$-direction:
\begin{gather}
t^{-1}_{j+r}\Lambda t_{j+r}\simeq t_j^{-1}\Bigl[\Lambda+ra[L^x_j,\Lambda]+\frac{(ra)^2}{2}([\partial_xL^x_j,\Lambda]\notag\\
+[L^x_j,[L^x_j,\Lambda]])+\dots\Bigr]t_j .
\label{eq:Qj:ra}
\end{gather}
Then, using sublattice representation for the matrix $\beta^{(0)}_{jj'}$, we find
\begin{align}
\tilde Q_j  & \! =\! \sum_{s=0,1}\sum_n \beta^{(0)}_{j,j+2n+s}\tilde{Q}_{j+2n+s}^{(0)} \simeq \frac{\textsf{N}}{2v} t_j^{-1}\Bigl[A_0 \Lambda +a(-1)^j \notag \\
 \times & A_1[L^x_j,\Lambda] +  a^2\frac{A_2}{2}([\partial_xL^x_j,\Lambda]+[L^x_j,[L^x_j,\Lambda]])\Bigr] t_j .
\label{Q_exp_general}
\end{align}
Here we introduce the following parameters ($m=0,1,2$)
\begin{equation}
 A_m= \sum_n \Bigl [(2n)^m \beta^{(0)}_{00}(-n)+(2n+1)^m\beta^{(0)}_{01}(-n)\Bigr ] .
\end{equation}
The coefficients $A_m$ can be written in a more compact form in terms of the function
\begin{gather}
\underline{\beta}(k)=\sum_{j'}\beta^{(0)}_{2j,2j+j'}e^{ikj'}
=\sum_{n}\beta^{(0)}_{00}(-n)e^{ik(2n)}\notag \\ +\sum_{n}\beta^{(0)}_{01}(-n)e^{ik(2n+1)} .
\end{gather}
We note that $\underline{\beta}(k)$ coincides with the Fourier transform introduced in the Toeplitz case, cf. Eq. \eqref{eq:FT:T}. Then, we obtain
\begin{equation}
A_0 = \underline{\beta}(0), \quad A_1= -i \underline{\beta}'(0), \quad A_2= -{\underline{\beta}''(0)} .  
\end{equation}

In order the expansion in $ra$ in Eq. \eqref{eq:Qj:ra} to be controlled we require the gradients to be sufficiently small. We assume that $\beta_{ss'}^{(0)}(n)$ decays exponentially on a scale $n_0$. Therefore, the matrix field $t$ must change on length scales greater than the scale $a n_0$. After carrying out the gradient expansion in a standard manner, we arrive at an effective action:
\begin{gather}
\mathcal S\simeq-\sum_j\int dy\Bigl\{\pi_y\tr[L_j^{(y)},\Lambda]^2+\pi_x\tr[L_j^{(x)},\Lambda]^2\notag\\
+ \pi_{H} \tr\,\Lambda[L_j^{(x)},L_j^{(y)}]- \frac{\textsf{N}}{2v}\tr (\tilde{\varepsilon}+i\tilde{b}_j) q_j \notag \\
- \frac{\textsf{N}\tau}{8 v}\tr[ t_j \tilde{b}_j t_j^{-1}, \Lambda]^2-(-1)^j\frac{\textsf{N}}{2}\tr\Lambda L_j^{(y)}\\
+
\frac{i \textsf{N}\tau}{4v} v_j \tr [t_j \tilde{b}_j t_j^{-1},\Lambda][L_j^{y},\Lambda] 
\Bigr\} ,
\label{eq:act:1112}
\end{gather}
where coefficients are given by 
\begin{gather}
\pi_x=-\frac{a^2 \underline{\beta}''(0)}{16\tau^2\underline{\beta}^2(0)}\left (1{-}\frac{2[\underline{\beta}'(0)]^2}{\underline{\beta}''(0)\underline{\beta}(0)}\right )
, \quad \pi_y=\frac{\textsf{N} v\tau}{8} ,\notag \\
\pi_{H} =i\frac{\textsf{N}^2\tau a}{v} \underline{\beta}'(0) .
\label{eq:act:1112:pi}
\end{gather}
Repeating the same steps of derivation as before, we again obtain the {\NLSM} action \eqref{final_action} with scattering time $\tau$ given by Eq. \eqref{eq:SP:1}, as well as with the following longitudinal and Hall conductivities 
\begin{gather}
\bar{g}=\textsf{N}\left [\frac{2|\underline{\beta}''(0)|}{\underline{\beta}(0)}\left (1{-}\frac{2[\underline{\beta}'(0)]^2}{\underline{\beta}''(0)\underline{\beta}(0)}\right )\right ]^{1/2} ,\notag \\ \bar{g}_H=\textsf{N}\Bigl ( 1+\frac{2i\underline{\beta}'(0)}{\underline{\beta}(0)}\Bigr )\label{eq:gg:g:gH}
\end{gather}
We note that in the symmetric case $\underline{\beta}'(0)\equiv 0$ and Eq.~\eqref{eq:gg:g:gH} reproduces Eq.~\eqref{eq:ff:g:gH}. For the nearest-neighbor tunneling, 
\begin{equation}
\underline{\beta}(k)=(3m/2)+\lambda_+e^{-ik}+\lambda_-e^{ik}, 
\label{eq:beta:NNT}
\end{equation}
such that Eq. \eqref{eq:gg:g:gH} reproduces the results \eqref{final_g} and \eqref{final_gH}. Therefore, all previously discussed limits can be seen as special cases of the general expression \eqref{eq:gg:g:gH}.

It is instructive to introduce two parameters $\epsilon=2i\underline{\beta}'(0)/\underline{\beta}(0)$ and $\phi=2|\underline{\beta}''(0)|/\underline{\beta}(0)$. Then, the expressions \eqref{eq:gg:g:gH} can be rewritten as follows, 
\begin{equation}
\bar{g}=\textsf{N}\sqrt{\phi-\epsilon^2},\qquad \bar{g}_H=\textsf{N}(1-\epsilon) . 
\label{eq:gg:g:gH:r}
\end{equation}

By choosing the function $\underline{\beta}(k)$, one can change the parameters $\phi$ and $\epsilon$, and, thus, control the starting point for the two-parameter renormalization group flow governed by the {\NLSM} action. Parameter $\epsilon$ is responsible for the deviation of the spin Hall conductance from the integer value. Parameter $\phi$ controls the bare value of the spin conductance. 
Provided the function $\underline{\beta}(k)$ satisfies the inequality $\re\underline{\beta}(k)>0$,~\footnote{We note that the inequality $\re\underline{\beta}(k)>0$ guarantees that $|\underline{\beta}(k)|>0$ and, consequently, the matrix $\beta^{(0)}$ has no zero eigenvalues. However, the absence of zero eigenvalues is possible even if $\re\underline{\beta}(k)<0$.} one can prove that the parameters $\phi$ and $\epsilon$ satisfy the following inequalities (see Appendix \ref{App_Det2}):
\begin{equation}
   \phi >  \epsilon^2 .
\end{equation}
This inequality guarantees that the longitudinal spin conductance is well-defined for $\re \underline{\beta}(k)>0$. We note that the necessary condition, $|\underline{\beta}(k)|>0$, for the absence of zero eigenvalues of the matrix $\beta^{(0)}$, does not guarantee that $\phi>\epsilon^2$ (see Appendix \ref{App_Det2}). We note that the regime $\phi<\epsilon^2$ is realized for nearest-neighbor tunneling (see Eq. \eqref{eq:beta:NNT}), provided that
$3m/(2(\lambda_++\lambda_-))<2(\lambda_+-\lambda_-)^2/(\lambda_++\lambda_-)^2-1$. This condition requires a relatively strong anisotropy in the tunneling between even and odd links, namely $|\lambda_+-\lambda_-|/(\lambda_++\lambda_-)>1/\sqrt{2}$.
The case when $\phi<\epsilon^2$ requires a separate analysis. In particular, for $\pi_x<0$, the saddle point given in Eq. \eqref{eq:sspp:11} does not correspond to a minimum of the action. However, a detailed investigation of this regime lies beyond the scope of the present manuscript.

\section{The effect of the triplet modes \label{Sec5}}

So far, we have ignored the triplet matrix fields $\tilde{Q}^{(c)}_j$, with $c=1,2,3$, in the action \eqref{action_d=g}. In this section, we discuss their effect on the {\NLSM} action. First, we assume that these triplet modes are massive on the level of the Gaussian action \eqref{eq:gauss:triple}. Second, we discuss the situation where the Gaussian action \eqref{eq:gauss:triple} may have some zero modes.

\subsection{Corrections to the {\NLSM} due to the presence of the triplet modes}

Under the assumption that the Gaussian action \eqref{eq:gauss:triple} has no zero modes, our aim is to integrate out the triplet modes and find corrections to the {\NLSM} action. Using Eq. \eqref{eq:gauss:triple}, we find the following propagator for the $\tilde{Q}^{(c)}_j$ modes:
\begin{gather}
  \langle 
  \spp 
  s_s [\tilde{Q}^{(c)}_j(y)]^{\alpha\beta}_{\varepsilon_1\varepsilon_2}
  \spp s_{s'} [\tilde{Q}^{(c')}_{j'}(y')]^{\mu\nu}_{\varepsilon_4\varepsilon_3}\rangle_0
  =  
  2 \delta(y-y') \notag \\
  \times \delta_{ss'} \Bigl [ \delta^{\alpha\nu}\delta^{\beta\mu} \delta_{\varepsilon_1\varepsilon_3}\delta_{\varepsilon_2\varepsilon_4}
  - \textsf{v}_s \delta^{\alpha\mu}\delta^{\beta\nu}\delta_{\varepsilon_1,-\varepsilon_4}\delta_{\varepsilon_2,-\varepsilon_3} \Bigr ]\notag \\
  \times \delta^{cc'}  \Bigl [ (\theta(\varepsilon_1)\theta(\varepsilon_2)+\theta(-\varepsilon_1)\theta(-\varepsilon_2)) [(\beta^{(c)})^{-1}]_{jj'}
  \notag \\
  +(\theta(\varepsilon_1)\theta(-\varepsilon_2)+\theta(-\varepsilon_1)\theta(\varepsilon_2)) [(\tilde{\beta}^{(c)})^{-1}]_{jj'} .
\Bigr ]
\label{eq:prop:Qaa}
\end{gather}
Here, we introduce the parameter $\textsf{v}_s=-\spp s_s s_2 s_s^Ts_2/2$, and $\theta(\varepsilon)$ denotes the Heaviside step function. Using Eq.~\eqref{eq:prop:Qaa}, one can derive the contraction rule:
\begin{gather}
\int dy' \langle \tr A(y',y) \tilde{Q}^{(c)}_j(y) B(y,y')  \tilde{Q}^{(c)}_{j'}(y') \rangle_0 =  (\check{\beta}^{(c)}_+)^{-1}_{jj'}
\notag \\ \times 
 \Bigl [ \tr A(y,y) \tr B(y,y) {-} \tr A(y,y) C B^T(y,y) C\Bigr ] 
 {+} (\check{\beta}^{(c)}_-)^{-1}_{jj'}\notag \\ \times 
 \Bigl [ \tr \Lambda A(y,y) \tr \Lambda B(y,y) + \tr \Lambda A(y,y)\Lambda  C B^T(y,y) C\Bigr ] ,
 \label{eq:rel:rel:Qa}
\end{gather}
where $(\check{\beta}^{(c)}_\pm)^{-1}=[({\beta}^{(c)})^{-1}\pm(\tilde{\beta}^{(c)})^{-1}]/2$. Next, we use the relation \eqref{eq:rel:rel:Qa} to compute the term that obtained after the second-order expansion of $\tr\ln$ in  $\tilde{Q}^{(c)}_{j'}$. We note that such an expansion must be performed carefully in order to preserve the invariance of the action 
\eqref{action_d=g} with respect to global rotations of the $Q_{j}^{(0)}$ matrix. Therefore, we proceed as follows: (i) we perform a rotation $t_j$ inside the $\tr\ln$ operator, and then (ii) we change variables from $\tilde{Q}^{(c)}_{j'}$ to $t_{j'}^{-1}\tilde{Q}^{(c)}_{j'} t_{j'}$ in the action \eqref{action_d=g}. Overall, we expand the $\tr\ln$ operator to second order in the quantity $t_j t_{j'}^{-1}\tilde{Q}^{(c)}_{j'} t_{j'} t_j^{-1}$. The result reads
\begin{gather}
\frac{\textsf{N}}{2} \int \!dy dy'\!\! \sum_{jj_1j_2,c}\beta^{(c)}_{jj_1}\beta^{(c)}_{jj_2} 
\tr \langle t_j(y) t_{j_1}^{-1}(y) \tilde{Q}^{(c)}_{j_1}(y)t_{j_1}(y) t_j^{-1}(y) \notag \\
\times  \hat{G}_j(y,y') t_j(y')  t^{-1}_{j_2}(y')\tilde{Q}^{(c)}_{j_2}(y') t_{j_2}(y') t_j^{-1}(y') \hat{G}_j(y',y)  \rangle_0,        
\label{eq:expan:Qa:0}
\end{gather}
where we introduce the inverse Green's function 
\begin{align}
\hat{G}_j^{-1} & = \underline{G}_j^{-1}+i v_j L_j^{(y)} +i\frac{\textsf{N}}{2v} \sum_{j'} \beta^{(0)}_{jj'} t_j t^{-1}_{j'}[\Lambda, t_{j'}]t_j^{-1} \notag \\
& \simeq \underline{G}_j^{-1} +i v_j L_j^{(y)} + i (-1)^j a\frac{\textsf{N} A_1}{2v} [L_j^{(x)},\Lambda] 
\notag \\
& + i a^2 \frac{\textsf{N} A_2}{4v}\Bigl([\partial_x L_j^{(x)},\Lambda]+[L_j^{(x)},[L_j^{(x)},\Lambda]] \Bigr ) .
\end{align}
Next, using the contraction rule \eqref{eq:rel:rel:Qa}, we obtain from Eq.~\eqref{eq:expan:Qa:0} the following result:
\begin{widetext}
\begin{gather} 
-\frac{\textsf{N}}{2}\int dy \sum_{j,j_1,j_2,c}
\beta^{(c)}_{jj_1} \beta^{(c)}_{jj_2} 
\Biggl \{  (\check{\beta}_+^{(c)})^{-1}_{j_1j_2} \Bigl [ \Bigl (\tr t_{j_1}(y) t_j^{-1}(y) \hat{G}_j(y,y) t_j(y)  t^{-1}_{j_2}(y)\Bigr )^2  +\tr \Bigl (t_{j_1}(y) t_j^{-1}(y) \hat{G}_j(y,y) t_j(y)  t^{-1}_{j_2}(y)\Bigr )^2 \Bigr ] \notag \\+  (\check{\beta}_-^{(c)})^{-1}_{j_1j_2} \Bigl [\Bigl (\tr \Lambda t_{j_1}(y) t_j^{-1}(y) \hat{G}_j(y,y) t_j(y)  t^{-1}_{j_2}(y)\Bigr )^2  -\tr \Bigl (\Lambda t_{j_1}(y) t_j^{-1}(y) \hat{G}_j(y,y) t_j(y)  t^{-1}_{j_2}(y)\Bigr )^2\Bigr ]\Bigr \} .
\label{eq:expan:Qa}
\end{gather}
\end{widetext}
Here, we use the fact that $t_j^T C= Ct_j^{-1}$ and 
$C \hat{G}_j^T(y,y) C=-\hat{G}_j(y,y)$. The term \eqref{eq:expan:Qa} is an additional contribution to the {\NLSM} action. It produces corrections to the gradient terms in the action \eqref{eq:act:1112}. In particular, we find the following corrections (cf. Eq.~\eqref{eq:act:1112:pi})
\begin{equation}
\delta \pi_y = \frac{\textsf{N}}{8} \sum_{c=1}^3 \tau_j^2 \beta^{(c)}_{jj},\quad 
\delta \pi_H = \frac{\textsf{N}^2 A_1 a}{2v^2} \sum_{c=1}^3 \tau_j^2 \beta^{(c)}_{jj}, 
\label{eq:act:1112:pi:c1}
\end{equation}
and
\begin{align}
\delta \pi_x = &  \frac{\textsf{N} a^2}{16 v^2} \sum_{c=1}^3 \sum_{j_1,j_2} 
(2j-j_1-j_2)^2\beta^{(c)}_{jj_1} (\check{\beta}^{(c)}_-)^{-1}_{j_1j_2} \beta^{(c)}_{j_2j} \notag \\
& -\frac{\textsf{N}^3A_1^2a^2}{8 v^4} \sum_{c=1}^3 \tau_j^2 \beta^{(c)}_{jj} .
\label{eq:act:1112:pi:c2}
\end{align}
We note that the contribution to $\delta \pi_x$ in the first line of Eq. \eqref{eq:act:1112:pi:c2} comes from the expansion of the products $t_{j_1} t_j^{-1}$ in gradients in Eq. \eqref{eq:expan:Qa}. All the other contributions in Eqs. \eqref{eq:act:1112:pi:c1} and \eqref{eq:act:1112:pi:c2} are the result of the expansion of the Green's function $\hat{G}_j(y,y)$. 
We emphasize that the corrections \eqref{eq:act:1112:pi:c1} and \eqref{eq:act:1112:pi:c2} are smaller than the expressions \eqref{eq:act:1112:pi} by a factor of $1/\textsf{N}$ . Therefore, the role of the triplet modes  $\tilde{Q}^{(c)}_j$, with $c=1,2,3$, is similar to the massive modes in the singlet sector. Indeed, the massive modes in the singlet sector can be studied in a similar manner. Formally, one can use similar expressions with $\tilde{\beta}^{-1}$ replaced by zero. Also, there is an additional contribution due to the Jacobian \cite{Zirnbauer1997}. Since the singlet massive modes are similar to the massive modes in the case of the {\NLSM} for class A \cite{pruisken1984localization}, we do not discuss them in the present work. 
In addition, we note that the corrections \eqref{eq:act:1112:pi:c1} and \eqref{eq:act:1112:pi:c2} exist even in the case $\gamma_j=\Delta_j$. This is due to the diagonal part of the matrix $\beta^{(c)}$, cf. Eq. \eqref{eq:beta:def1}. We recall that for $\textsf{N}>1$ one cannot exclude the triplet modes $D^{(c)}_{j}$, where $c=1,2,3$, from the theory even for  $\Delta_j=\gamma_j$.

\subsection{Possible softening of the triplet modes}

In the previous section, we used Eq. \eqref{eq:prop:Qaa} to derive corrections to the {\NLSM} action due to the triplet modes. We assumed that the matrices $\beta^{(c)}_{jj'}$ and  $\tilde{\beta}^{(c)}_{jj'}$ are invertible, meaning they have no zero eigenvalues. This assumption is reasonable for the matrix $\beta^{(c)}_{jj'}$ as it is involved in the Hubbard-Stratonovich decoupling in Eq. \eqref{eq:dis:av:00}. 
If the matrix $\beta^{(c)}_{jj'}$ had zero eigenvalues, the Hubbard-Stratonovich decoupling would have to be modified. Below we assume that this is not the case. In contrast, the matrix $\tilde{\beta}^{(c)}_{jj'}$ is not directly involved in the  Hubbard-Stratonovich decoupling and, therefore, may occasionally have zero eigenvalues. Therefore, it is important to analyze this case in more detail.

According to Eq. \eqref{eq:gauss:triple}, the matrix $\tilde\beta^{(c)}_{jj'}$ determines the Gaussian action for the part of the matrix  $\tilde Q^{(c)}_j$ that anticommutes with  
$\Lambda$, $\tilde Q_{j,-}^{(c)}=\Lambda[\Lambda,\tilde Q^{(c)}_j]/2$. We diagonalize the symmetric matrix, $\tilde\beta^{(c)}$, by an orthogonal rotation, $R$, such that  $\tilde\beta^{(c)}_{jj'}=\sum_r R_{rj} \tilde{\beta}^{(c)}_r R_{rj'}$. We assume that $\tilde{\beta}^{(c_0)}_{r_0}\equiv 0$ for some indices $c_0$ and $r_0$. The corresponding zero mode is given by $Z_{r_0}^{(c_0)}=\sum_j R_{r_0j}\tilde Q_{j,-}^{(c_0)}$. 
To demonstrate that this mode is a soft mode rather than a true zero mode, we expand the action \eqref{action_d=g} to fourth order in $\tilde Q^{(c)}_j$ with $c=1,2,3$ (neglecting the singlet mode $\tilde Q^{(0)}_j$). The resulting contribution to the action is as follows
\begin{gather}
 \mathcal{S}_{\rm t}^{(4)} =-\frac {\textsf{N}}{16v}\sum_{j,j_{k}}\tau_j^3\sum_{c_k=1}^3\beta_{jj_1}^{(c_1)}\beta_{jj_2}^{(c_2)}\beta_{jj_3}^{(c_3)}\beta_{jj_4}^{(c_4)}  \Bigl (\delta_{c_1c_2}\delta_{c_3c_4}\notag \\
 -\delta_{c_1c_3}\delta_{c_2c_4}+\delta_{c_1c_4}\delta_{c_2c_3}\Bigr ) \int dy
 \tr \Bigl[\tilde Q_{j_1}^{(c_1)}\tilde Q_{j_2}^{(c_2)}\tilde Q_{j_3}^{(c_3)}\tilde Q_{j_4}^{(c_4)}
 \notag\\
 -4 \Lambda \tilde  Q_{j_1}^{(c_1)}\Lambda  \tilde Q_{j_2}^{(c_2)}\tilde Q_{j_3}^{(c_3)}\tilde Q_{j_4}^{(c_4)}
 -2 \Lambda  \tilde  Q_{j_1}^{(c_1)}\tilde Q_{j_2}^{(c_2)}\Lambda   \tilde Q_{j_3}^{(c_3)}\tilde Q_{j_4}^{(c_4)}
  \notag\\
 +5\Lambda  \tilde  Q_{j_1}^{(c_1)}\Lambda \tilde Q_{j_2}^{(c_2)}\Lambda   \tilde Q_{j_3}^{(c_3)}\Lambda \tilde Q_{j_4}^{(c_4)})\Bigr] .
\end{gather}
Here, we neglect all spatial derivatives with respect to the coordinate $y$. Projecting the above lengthy expression onto the mode $Z_{r_0}^{(c_0)}$, we obtain
\begin{gather}
 \mathcal{S}_{\rm t}^{(4)} \to -\frac {\textsf{N}}{2v}\sum_j\tau_j^3 \left (\sum_{j'}\beta_{jj'}^{(c_0)} R_{j'r_0}\right )^4\int dy
 \tr (Z_{r_0}^{(c_0)})^4 .
 \label{eq:Z4th}
\end{gather}
Since $R_{j'r_0}$ cannot simultaneously be a zero eigenvector for both matrices $\tilde\beta^{(c)}_{jj'}$ and $\beta^{(c)}_{jj'}$, we conclude that there is a quartic term for the mode $Z_{r_0}^{(c_0)}$; i.e., this mode is a soft mode rather than a true zero mode.  
The quartic action \eqref{eq:Z4th} indicates that the contribution of the soft mode $Z_{r_0}^{(c_0)}$  to the correlation function $\langle \tilde Q^{(c_0)}_{j,-}\tilde Q^{(c_0)}_{j',-}\rangle$ is enhanced by the factor $\textsf{N} \sqrt{m L_y/v}$, where $L_y$ is the system size in the $y$ direction. This enhancement may result in corrections  similar to those in Eqs.~\eqref{eq:act:1112:pi:c1}--\eqref{eq:act:1112:pi:c2}, which are not small in the parameter $1/\textsf{N}$. However, a detailed analysis of this case is beyond the scope of the present work.

\section{Discussion and conclusions \label{Sec6}}

We demonstrated that the continuum limit of the quantum network model at large $\textsf{N}$ is equivalent to the {\NLSM} in the weak coupling regime. Interestingly, for the general formulation of the model, the triplet sector of the theory does not decouple from the singlet sector. Since the form of the {\NLSM} is dictated solely by symmetry considerations, the inclusion of triplet modes does not modify the structure of the {\NLSM} action. However, they have two important effects.
First, the triplet modes renormalize the bare value of the longitudinal spin Hall conductance in the {\NLSM}. In the large-$\textsf{N}$ limit, these corrections to $\bar{g}$ are suppressed as $1/\textsf{N}$, see Eq. \eqref{eq:act:1112:pi:c2}. Second, for certain choices of tunneling between links, the triplet modes may become massless at the Gaussian level, although they remain massive beyond the Gaussian approximation. This indicates that the triplet modes are anomalously soft and, as a consequence, modify the ultraviolet length scale of the {\NLSM}. Additionally, in this case, the correction to $\bar{g}$ produced by the triplet modes may not be small in $1/\textsf{N}$ and thus, it may also be significant. We note that such choices of tunneling between links should be avoided in numerical simulations of the quantum network model, as the required system sizes needed to observe universal effects may be increased compared to the standard scenario.

\begin{figure}[t]
\centerline{\includegraphics[width=0.85\columnwidth]{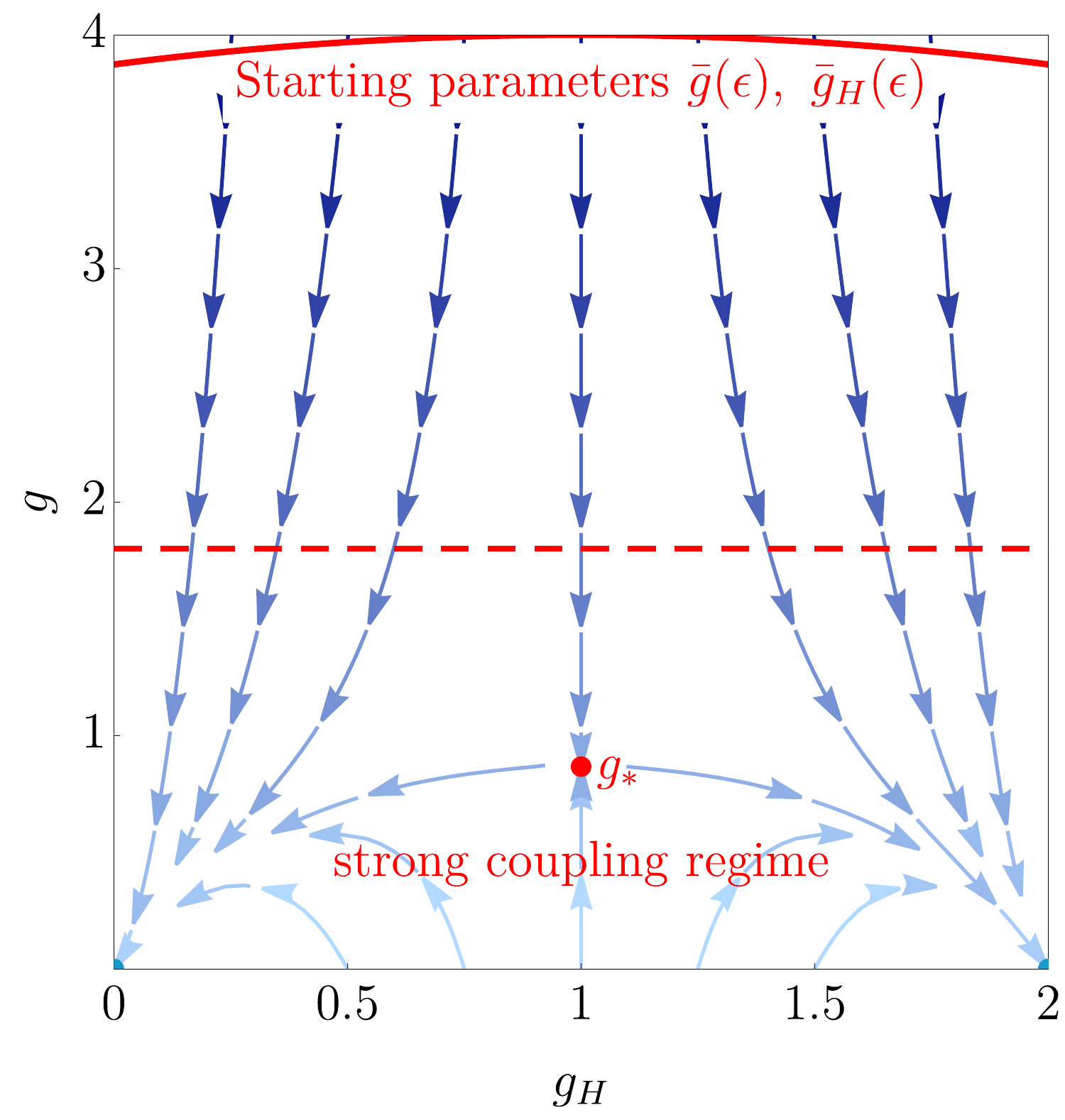}}
\caption{Sketch of the renormalization group flow for class C {\NLSM} (adopted from Ref. \cite{Parfenov2024}). Arrows indicate flow towards infrared. Red curve at the top demonstrates a change of the bare parameters $\bar{g}$ and $\bar{g}_H$ with $\epsilon$ at fixed $\phi$. }
\label{fig:RG_flow}
\end{figure}

The quantum network model considered in this work is a generalization of the model proposed for the {\sqHe} in Ref.~\cite{Senthil1999} and extended to a large number of channels $\textsf{N}$ per link. Although this model is similar to the quantum network model studied in Ref.~\cite{DHLee1998} for the {\iqHe}, there is a crucial difference in the bare conductances. In the case of the {\iqHe}, both bare conductances are controlled by a single parameter. In contrast, in the considered case of the {\sqHe}, the bare longitudinal spin conductance is governed by two independent parameters, one of which also enters the bare spin Hall conductance. This fact makes the considered quantum network model convenient    
for numerical studies of the {\sqHe} flow diagram (see Fig.~\ref{fig:RG_flow}), since one can change the bare values of the conductances independently.

Perhaps, the most striking result of our analysis is that the absence of zero eigenvalues in the tunneling matrix $\beta^{(0)}$, i.e., the inequality  $|\underline{\beta}(k)|>0$, does not guarantee 
that $\phi>\epsilon^2$ in Eq. \eqref{eq:gg:g:gH:r}. The case $\phi<\epsilon^2$ implies that the saddle point \eqref{eq:sspp:11} does not correspond to a minimum of the action. Since fulfillment of the inequality $\phi<\epsilon^2$ requires sufficient anisotropy in the tunneling between even and odd links, perhaps it implies that in such a situation the quantum network effectively  splits into pairs of neighboring links decoupled from each other. We leave a detailed analysis of this possibility for future work.

We also considered the effect of the Zeeman field acting on fermions in each link. As expected, we found that the Zeeman splitting leads to terms in the {\NLSM} action that break SU(2) symmetry, see Eq. \eqref{final_action}. These terms are relevant and responsible for the crossover from the {\sqHe} to the {\iqHe} \cite{Gruzberg1999,BhardwajKagalovsky2015,Parfenov2025}.

The {\NLSM} for class C has been derived as an effective description in the large $\textsf{N}$ limit of a system of $2\textsf{N}$ species of 2D Dirac fermions with class C symmetry, as described in Ref. \cite{Bernard2001}. However, for the case of isotropic disorder, it was found \cite{Bernard2001} that the bare value of the spin Hall conductance, $\bar{g}_H$, is equal to $2\textsf{N}$. This implies that {\NLSM} is on a line where all bulk states are localized, and the physical spin Hall conductance is also equal to $2\textsf{N}$. This situation contrasts with the quantum network model we consider in this work, where for odd $\textsf{N}$ and symmetric tunneling ($\epsilon=0$), the {\NLSM} lies on the critical line.

We emphasize the difference between the quantum network model we consider and the SU(2) Chalker-Coddington network model. The former has random tunneling amplitudes, while the latter has two fixed tunneling amplitudes. Their asymmetry determines the parameter $\epsilon$ in the bare value of the spin Hall conductance. Simultaneously, in the Chalker-Coddington model, the parameter $\phi$ (see Eq. \eqref{eq:gg:g:gH:r}) is always equal to $1$. Since the {\NLSM} for the Chalker-Coddington model is derived from a mapping to a spin chain, it is interesting to understand which type of spin chain corresponds to the {\NLSM} action derived in our work and what the meaning of $\phi$ is for the spin chain Hamiltonian.

Recently, there have been indications that the Chalker-Coddington network model may be too simplistic to accurately describe the critical behavior of the {\iqHe}. Numerical studies have shown that geometric disorder in the Chalker-Coddington model for the {\iqHe} leads to a modification of the critical exponent of the correlation length \cite{Sedrakyan2017,Sedrakyan2019,Sedrakyan2021,Dresselhaus2022,Topchyan2024}. In the case of the {\sqHe}, a modification of the critical exponents has recently been shown for random networks \cite{Gruzberg2026}. In this context, it would be interesting to extend the quantum network model considered in this work to include the effects of geometric disorder. One possible step in this direction would be to introduce finite correlation lengths along the $y$ direction for the tunneling amplitudes, see Eq. \eqref{eq:t:av}.

So far, the experimental realization of the {\sqHe} remains elusive. Originally, observation of the {\sqHe} was suggested in the $d_{x^2-y^2} + i d_{xy}$ topological superconducting state \cite{Volovik1997,Kagolovsky1999,Senthil1999}. However, to date, there is no experimental evidence of the existence of a  superconducting film with the $d_{x^2-y^2} + i d_{xy}$ order parameter symmetry. Recently, experiments on twisted Bi$_2$Sr$_2$CaCu$_2$O$_{8+x}$ bilayers have shown some indications  
of spontaneously breaking time-reversal symmetry at a $45^\circ$ twist angle \cite{Zhao-Science,Martini2023}. This observation is in agreement with theoretical
predictions for an emergent $d_{x^2-y^2} + i d_{xy}$ topological superconducting state in  twisted superconducting flakes~\cite{Volkov,Franz}. Another proposal to realize the {\sqHe} is based on the observation that  edge states at the interface between a two-dimensional electron gas in a perpendicular magnetic field and an $s$-wave superconductor are actually the {\sqHe} edge states \cite{Parfenov2026}. This observation suggests that the quantum network model discussed in our work could potentially be realized in a two-dimensional electron gas with an $s$-wave superconducting islands in the presence of a perpendicular magnetic field (see, e.g., Ref. \cite{Han2014}). 

Twisted Bi$_2$Sr$_2$CaCu$_2$O$_{8+x}$ bilayers with twist angles near $45^\circ$ also demonstrate the Josephson diode effect \cite{Zhao-Science}. As is known, the superconducting diode effect requires the breaking of time-reversal and inversion symmetries (see Ref. \cite{Nadeem2023} for a review). We note that 
the quantum network model we consider in this work does not have time-reversal symmetry or inversion symmetry for a given realization of random tunneling. After disorder averaging, the inversion symmetry is restored unless the Zeeman splitting is taken into account. In this case, the {\NLSM} action, see Eq. \eqref{final_action}, includes a term that  explicitly breaks  inversion symmetry.~\footnote{In chiral symmetry classes, the Berry-phase weak topological term takes a familiar form; for a review, see, for example, Ref.~\cite{Shindou2024}.} Therefore, it would be interesting to investigate non-reciprocal effects associated with broken inversion symmetry in the quantum network model. 

To summarize, in this study we examined a quantum network model with $\textsf{N}$
channels per chiral link, adhering to the symmetries of the spin quantum Hall effect. We established that the triplet and singlet sectors remain coupled, though triplet modes may influence the derived large-$\textsf{N}$ {\NLSM} only when they become ``soft'' under specific conditions. Calculations of bare conductances in the {\NLSM} revealed that the standard saddle-point approximation fails in the case of sufficiently large asymmetry in tunneling between even and odd links. Furthermore, the inclusion of a Zeeman field was shown to break not only the SU(2)
 symmetry but also the inversion symmetry within the {\NLSM} action.

\begin{acknowledgments}
We are grateful to Y. Fominov, I. Gruzberg, and P. Ostrovsky for very useful and inspiring discussions. This work was supported by the Russian Science Foundation under Grant No. 24-12-00357. The authors acknowledge personal support from the Foundation for the Advancement of Theoretical Physics and Mathematics ``BASIS''.
\end{acknowledgments}

\appendix


\section{Details of evaluation for the ballistic RG\label{App:BRG}}

Here we present the evaluation of the average appearing in Eq. \eqref{eq:RG_no_avg}. Since this expression contains four fermionic fields, several types of Wick contractions are possible. However, in the large-$\textsf{N}$ limit, only contractions that are leading in $\textsf{N}$ need to be considered. These are the contractions between two $D$ matrices. Thus, we write
\begin{gather}
\left \langle 
 [D^{(a)}_{j'}(y)]_{\varepsilon_m,\beta,\sigma^\prime;\varepsilon_n,\alpha,\sigma} 
  [D^{(a)}_{j_1}(y')]_{\varepsilon_{m_1},\beta_1,\sigma_1^\prime;\varepsilon_{n_1},\alpha_1,\sigma_1}
  \right \rangle_0 \notag \\
  =
 \textcircled{1}+ \textcircled{2} .
  \end{gather}
The first contribution corresponds to a normal-type contraction:
  \begin{align}
 \textcircled{1} = & -\sum_{s,s',s_1,s_1',f,f_1}
 \bigl \langle 
  \Xi_{\varepsilon_m\beta j'f\sigma's'}(y)
  \bar{\Xi}_{\varepsilon_{n_1}\alpha_1 j_1f_1\sigma_1s_1}(y')
  \bigr\rangle_0
  \notag \\
  & \times  (s_a)_{ss'} (s_b)_{s_1s_1'} 
\bigl \langle 
  \Xi_{\varepsilon_{m_1}\beta_1 j_1f_1\sigma_1's_1'}(y')
  \bar{\Xi}_{\varepsilon_{n}\alpha j'f\sigma s}(y)
  \bigr \rangle_0 .
  \end{align}
The second contribution is the anomalous-type contraction:
  \begin{align}
  \textcircled{2} = & \sum_{s,s',s_1,s_1',f,f_1}
  \bigl \langle 
  \Xi_{\varepsilon_m\beta j'f\sigma's'}(y)
  \Xi_{\varepsilon_{m_1}\beta_1 j_1f_1\sigma_1's_1'}(y')
  \bigr\rangle_0\notag \\
  & \times  (s_a)_{ss'} (s_b)_{s_1s_1'}
  \bigl \langle 
  \bar{\Xi}_{\varepsilon_{n_1}\alpha_1 j_1f_1\sigma_1s_1}(y')
  \bar{\Xi}_{\varepsilon_{n}\alpha j'f\sigma s}(y)
  \bigr \rangle_0
  .
\end{align}  
Evaluating the normal contraction, we obtain
\begin{align}
\textcircled{1}=& \frac{\textsf{N}}{4}\delta_{j_1j'}\delta_{\varepsilon_{n_1}\varepsilon_{m}}   \delta_{\varepsilon_{m_1}\varepsilon_{n}} \delta_{\alpha_1\beta}\delta_{\beta_1\alpha}
\delta_{\sigma'\sigma_1}\delta_{\sigma_1'\sigma}\notag \\
& \times
\spp \Bigl [ s_a g_{\varepsilon_m,j_1}(y,y') s_b g_{\varepsilon_n,j_1}(y',y) \Bigr ].
\end{align}
Using the symmetry relation \eqref{eq:rel:Xi:Xi}, the anomalous contribution can be written as
\begin{align}
\textcircled{2}=& -\frac{\textsf{N}}{4}\delta_{j_1j'}\delta_{\varepsilon_{n_1},-\varepsilon_{n}}   \delta_{\varepsilon_{m_1},-\varepsilon_{m}} \delta_{\alpha_1\alpha}\delta_{\beta_1\beta}
(\sigma_2)_{\sigma'\sigma_1'}(\sigma_2)_{\sigma_1\sigma}\notag \\
& \times 
\spp \Bigl [ s_a g_{\varepsilon_m,j_1}(y,y') s_2s^T_bs_2 g_{\varepsilon_n,j_1}(y',y) \Bigr ] .
\end{align}
Here, $g_{\varepsilon_m,j}(y,y')$ denotes the single-particle Green's function in $d=1$. In the Nambu space, 
it takes the following form:
\begin{widetext}
\begin{equation}
    g_{\varepsilon_m,j}(y,y') =\begin{pmatrix}
    \langle y|(i\varepsilon_m+i v_j \partial_y)^{-1}|y'\rangle & 0  
    \\
    0 & -  \langle y'|(-i\varepsilon_m+i v_j \partial_y)^{-1}|y\rangle
    \end{pmatrix}
    = s_0 \int \frac{dq}{2\pi} \frac{e^{-iq(y-y')}}{i\varepsilon_m - v_j q} . 
\end{equation}
\end{widetext}
Combining the two contributions, we find that the leading term in the large-$\textsf{N}$ limit becomes
\begin{align}
 \textcircled{1}& +\textcircled{2}=  \frac{\textsf{N}}{2}\delta_{j_1j'}\delta_{ab}\Bigl[ \delta_{\varepsilon_{n_1}\varepsilon_{m}}   \delta_{\varepsilon_{m_1}\varepsilon_{n}} \delta_{\alpha_1\beta}\delta_{\beta_1\alpha}
\delta_{\sigma'\sigma_1}\delta_{\sigma_1'\sigma}\notag \\
& {+} (1{-}2\delta_{a0})\delta_{\varepsilon_{n_1},{-}\varepsilon_{n}}   \delta_{\varepsilon_{m_1},{-}\varepsilon_{m}} \delta_{\alpha_1\alpha}\delta_{\beta_1\beta}
(\sigma_2)_{\sigma'\sigma_1'}(\sigma_2)_{\sigma_1\sigma}\Bigr ]\notag \\
& \times
\int \frac{dqdk}{(2\pi)^2} \frac{e^{-ik(y-y')}}{(i\varepsilon_m - v_{j_1} q)(i\varepsilon_n - v_{j_1} (q+k))}  
\end{align}
In the low-energy limit, this expression yields local terms that renormalize the action in accordance with Eq. \eqref{eq:corr:ballistic}.

\section{Evaluation of determinants of matrices $\beta^{(a)}$ and $\tilde{\beta}^{(a)}$  \label{App_Det1}}

Throughout the derivation of the {\NLSM} we assumed that the matrices $\beta^{a}$ do not possess zero eigenvalues, since such eigenvalues would generate additional zero modes. This assumption is equivalent to requiring that $\det\beta^{(a)}\neq 0$. Here we analyze these conditions in more detail. 

\subsection{The case of the nearest neighbouring tunneling}

We begin with the tridiagonal matrix $\beta^{(a)}$ introduced in Eq. \eqref{eq:beta:def1}. For the singlet sector, the determinant of the $M\times M$ matrix $\beta^{(0)}$ can be written as
\begin{equation}
    \det\beta^{(0)}_M=d_M=\begin{vmatrix}
     \frac32m & \lambda_+ & 0 & 0 & \dots \\
     \lambda_+ & \frac32m & \lambda_- & 0 & \dots \\
     0 & \lambda_- & \frac32m & \lambda_+ & \dots \\
     0 & 0 & \lambda_+ & \frac32m & \dots \\
     \dots & \dots & \dots & \dots & \dots 
    \end{vmatrix}.
\end{equation}
Using standard properties of determinants, we find the following recursion relation
\begin{equation}
d_{n+2}=(9m^2/4-\lambda_+^2-\lambda_-^2) d_{n}-\lambda_+^2\lambda_-^2d_{n-2}.
\end{equation}
Since this is a linear recurrence, its solution can be expressed in terms of the roots $x_{\pm}$ of the characteristic equation:
\begin{equation}
x^2-\left(9m^2/4-\lambda_+^2-\lambda_-^2\right) x+\lambda_+^2\lambda_-^2=0 ,
\end{equation}
which has two roots:
\begin{equation}
x_\pm{=}\frac14\left(\sqrt{\frac94m^2{-}(\lambda_+{+}\lambda_-)^2}{\pm}\sqrt{\frac94m^2{-}(\lambda_+{-}\lambda_-)^2}\right)^2 .
\end{equation}
Then the odd and even determinants are given by
\begin{gather}
d_{2k-1}{=}\frac{d_2{-}x_-}{x_+{-}x_-} (\lambda_+^2{+}x_+)x_+^{k-1}{+}\frac{x_+{-}d_2}{x_+{-}x_-} (\lambda_+^2{+}x_-) x_-^{k-1} , \notag \\
 d_{2k}=\frac{d_2-x_-}{x_+-x_-} x_+^k+\frac{x_+-d_2}{x_+-x_-} x_-^k ,\label{finite_dets} 
 \end{gather}
with the initial conditions  $d_1=3m/2$ and $d_2=\frac94m^2-\lambda_+^2$. The conditions for the odd and even determinants to vanish are therefore
\begin{equation}
\left (\frac{x_-}{x_+}\right )^{k-1} {=} \frac{d_2{-}x_-}{d_2{-}x_+}\cdot \frac{\lambda_+^2{+}x_+}{\lambda_+^2{+}x_-}, \quad \left (\frac{x_-}{x_+}\right )^k= \frac{d_2{-}x_-}{d_2{-}x_+} .\label{cond_d2k:cond_d2k+1}
\end{equation}
Several cases are possible:
\begin{itemize}
\item If $3m/2>\lambda_++\lambda_-$, the roots are real and positive, with $d_2>x_+>x_->0$. In this regime, Eqs. \eqref{cond_d2k:cond_d2k+1} cannot be satisfied.

\item If  $3m/2<|\lambda_+-\lambda_-|$ both roots are negative, $x_-<x_+<0$. In this case, Eqs. \eqref{cond_d2k:cond_d2k+1} may be satisfied for a finite matrix size $M$. However, in the infinite matrix limit, $M\to\infty$, the contribution proportional to $x_+^k$ vanishes, and the remaining term cannot be tuned to zero.
    
\item In the intermediate regime,  $|\lambda_+-\lambda_-|<3m/2<\lambda_++\lambda_-$, the roots are complex and can be written as $x_\pm = |x| e^{\pm i \phi}$. Then the condition $d_{2k}=0$ can be rewritten as $d_2 \sin (\phi k)=|x| \sin(\phi(k-1))$, and an analogous, slightly more complicated relation could be obtained for $d_{2k+1}=0$. Thus, zero value of the determinants are possible.
\end{itemize}


We  therefore conclude that, in order to exclude zero modes of $\beta^{(0)}$, it is sufficient to impose the condition $m>2(\lambda_++\lambda_-)/3=2(\gamma_++\gamma_-)+2(\Delta_++\Delta_-)/3$. In the limit of infinitely many links, $M\to \infty$, this condition can be relaxed by additionally allowing the region $m<2|\lambda_+-\lambda_-|/3$. 

For the spin-triplet sector, the analysis is identical, with the substitutions $3m/2\to m/2$ and $\lambda_\pm\to\gamma_\pm-\Delta_\pm$. The corresponding positivity condition would be $m>2|\gamma_+-\Delta_+|+2|\gamma_--\Delta_-|$. A sufficient set of conditions ensuring positivity of $\tilde\beta^{(a)}_{jj'}$ for $a=0,1,2,3$ is
\begin{equation}
\begin{split}
m &>2(\gamma_++\gamma_-)+\frac23(\Delta_++\Delta_-) ,\\
m &>2|\gamma_+-\Delta_+|+2|\gamma_--\Delta_-| .
\end{split}
\end{equation}
Under these conditions, both coefficients $\pi_x$ and $\pi_y$ in Eq. \eqref{eq:S:interm} are positive and the inequality $\phi>\epsilon^2$ holds. 

\subsection{Tunneling matrix of a general form\label{App_Det2}}

We now consider  a general tunneling matrix $\beta_{jk}^{(a)}$ corresponding to the function
\begin{equation}
\underline{\beta}(k) = \beta_0 + \sum_{n=1}^\infty \Bigl ( \beta_n^{(+)} e^{-i k n}+\beta_n^{(-)} e^{i k n} \Bigr )   ,
\end{equation}
with $\beta_n^{(\pm)}>0$ and $\beta_0>0$. It is convenient to introduce $\beta_n=\beta_{n}^{(+)}+\beta_{n}^{(-)}$ and $\delta_n=(\beta_{n}^{(+)}-\beta_{n}^{(-)})/(\beta_{n}^{(+)}+\beta_{n}^{(-)})$, so that $|\delta_n|\leqslant 1$. Let us assume that the diagonal element is sufficiently large:
\begin{equation}
\beta_0>\sum_{n=1}^\infty \beta_n=\sum_{n=1}^\infty (\beta_{n}^{(+)}+\beta_{n}^{(-)}).\label{eq:SDDC}
\end{equation}
Under this assumption, $\Re \underline{\beta}(k)>0$ for all $k$ and the matrix $\beta_{jj'}^{(a)}$ is positive definite. Simultaneously, the inequality $\phi>\epsilon^2 $, or equivalently $\pi_{x,y}>0$ is also satisfied. Indeed, the inequality $\phi>\epsilon^2 $ is equivalent to 
\begin{equation}
\left  (\beta_0+\sum_{n=1}^\infty \beta_n\right ) \sum_{n=1}^\infty n^2   \beta_n > 2 \left (\sum_{n=1}^\infty n \delta_n \beta_n\right )^2.
\end{equation}
To prove that this relation holds, we use the inequality
\begin{equation}
    \sum_{n=1}^\infty \beta_n  \sum_{n=1}^\infty n^2   \beta_n \geqslant \left (\sum_{n=1}^\infty n \beta_n\right )^2,
\end{equation}
that follows from the Cauchy-Schwarz inequality $\bm{a}^2\bm{b}^2\geqslant (\bm{a}\bm{b})^2$ upon setting $a_n=\sqrt{\beta_n}$ and $b_n= n \sqrt{\beta_n}$. Therefore, we find
\begin{gather}
\left  (\beta_0+\sum_{n=1}^\infty \beta_n\right ) \sum_{n=1}^\infty n^2   \beta_n > 
2 \sum_{n=1}^\infty \beta_n \sum_{n=1}^\infty n^2   \beta_n\notag \\
\geqslant 
2 \left (\sum_{n=1}^\infty n \beta_n\right )^2\geqslant 
2 \left (\sum_{n=1}^\infty n \delta_n \beta_n\right )^2 .
\end{gather}
Hence, the condition \eqref{eq:SDDC} is sufficient to guarantee both the absence of zero eigenvalues and the positivity of the {\NLSM} coefficients. 

However, the condition \eqref{eq:SDDC} is not necessary. 
In order to demonstrate this fact, it is enough to consider the following example: $\underline{\beta}(k)=\beta_0+\beta_1 \cos k- i \delta_1 \beta_1 \sin k$ with $1/\sqrt{2}<|\delta_1|<1$. Then except for the special case $\beta_0=\beta_1$, one has $|\underline{\beta}(k)|>0$. Nevertheless, for $\beta_0/\beta_1<2\delta_1^2-1<1$, one finds $\phi<\epsilon^2$. 

\bibliography{literature_classC_top}

\end{document}